\pdfoutput=1

\documentclass[UKenglish, thm-restate]{lipics-v2021}
\title{Decomposing Permutation Automata}

\titlerunning{}%optional, please use if title is longer than one line

\author{Isma\"el Jecker}{Institute of Science and Technology, Austria}{ismael.jecker@ist.ac.at}{}{
	Marie Sk{\l}odowska-Curie Grant Agreement No. 754411
}%TODO mandatory, please use full name; only 1 author per \author macro; first two parameters are mandatory, other parameters can be empty. Please provide at least the name of the affiliation and the country. The full address is optional

\author{Nicolas Mazzocchi}{IMDEA Software Institute, Madrid, Spain}{nicolas.mazzocchi@imdea.org}{}{}%{[funding]}{[funding]}

\author{Petra Wolf}{Universit{\"a}t Trier, Fachbereich IV, Informatikwissenschaften, Germany \and \url{https://www.wolfp.net/}}{wolfp@informatik.uni-trier.de}{https://orcid.org/0000-0003-3097-3906}{DFG project  FE 560/9-1}

\authorrunning{I.~Jecker and N.~Mazzocchi and P.~Wolf}%TODO mandatory. First: Use abbreviated first/middle names. Second (only in severe cases): Use first author plus 'et al.'

\Copyright{Isma\"el~Jecker and Nicolas~Mazzocchi and Petra~Wolf}%TODO mandatory, please use full first names. LIPIcs license is "CC-BY";  http://creativecommons.org/licenses/by/3.0/


%TODO mandatory; please add comma-separated list of keywords

\category{}%optional, e.g. invited paper

\relatedversion{}%optional, e.g. full version hosted on arXiv, HAL, or other respository/website
\supplement{}%optional, e.g. related research data, source code, ... hosted on a repository like zenodo, figshare, GitHub, ...

\nolinenumbers %uncomment to disable line numbering

\hideLIPIcs  %uncomment to remove references to LIPIcs series (logo, DOI, ...), e.g. when preparing a pre-final version to be uploaded to arXiv or another public repository

\usepackage[vlined,linesnumberedhidden,resetcount]{algorithm2e} % algo
\usepackage{xspace}
\usepackage{bbold} % N, Z
\usepackage{color} % for debug only
\usepackage{mathtools}
\usepackage{tikz}
\usetikzlibrary{decorations.pathreplacing,calc}
\usepackage[bibliography=common]{apxproof}
\usepackage{xpatch}
\usepackage{apptools}

\tikzset{thick,>=stealth}
\tikzstyle{loop above right}=[out=60,in=30, min distance=5mm, looseness=8]
\tikzstyle{loop above left}=[out=150,in=120, min distance=5mm, looseness=8]
\tikzstyle{loop below left}=[out=-120,in=-150, min distance=5mm, looseness=8]
\tikzstyle{loop below right}=[out=-30,in=-60, min distance=5mm, looseness=8]
\tikzstyle{state}=[minimum size=15pt, circle,draw]
\tikzstyle{threestate}=[minimum width=25pt,minimum height=15pt,rectangle,rounded corners,draw]
\tikzstyle{format} = [text depth = 0.5cm, text height = 0.75cm]
\tikzstyle{transition}=[->,thick,>=stealth,shorten >=1pt,shorten <=1pt]
\tikzstyle{trans}=[->,>=stealth,shorten >=1pt,shorten <=1pt]
\tikzstyle{bitransition}=[<->,thick,>=stealth,shorten >=1pt,shorten <=1pt]
\tikzstyle{space}=[fill=white,inner sep=2pt]
\tikzstyle{lowlab}=[above,  yshift=-.075cm]

\newcommand{\showpictures}[1]{#1}

\newcommand{\showalgo}[1]{#1}

\newcommand{\drawSquare}[1]{
	\foreach \xx [evaluate=\xx as \x using {\xx-1}] in {2,...,#1}{
		\foreach \yy [evaluate=\yy as \y using {\yy-2}] in {2,...,#1}{
			\fill (\x,\y) circle (.2cm);
			\draw[<-, yshift=-.2cm] (\x,\y+1) -- ++(0, -.6); %^
			\draw[<-, dotted, yshift=-.2cm] (\x,\y) -- ++(0, -.6); %v
			\draw[<-, xshift=-.2cm] (\x,\y) -- ++(-.6, 0); %<
			\draw[<-, dotted, xshift=-.2cm] (\x+1,\y) -- ++(-.6, 0); %>
		}
	}
	\foreach \xx [evaluate=\xx as \x using {\xx-2}] in {2,...,#1}{
		\foreach \yy [evaluate=\yy as \y using {\yy-1}] in {2,...,#1}{
			\fill (\x,\y) circle (.2cm);
			\draw[<-, dotted, yshift=-.2cm] (\x,\y+1) -- ++(0, -.6); %^
			\draw[<-, yshift=-.2cm] (\x,\y) -- ++(0, -.6); %v
			\draw[<-, dotted, xshift=-.2cm] (\x,\y) -- ++(-.6, 0); %<
			\draw[<-, xshift=-.2cm] (\x+1,\y) -- ++(-.6, 0); %>
		}
	}
	\foreach \xx [evaluate=\xx as \x using {\xx-1}] in {1, #1}{
		\fill (\x,\x) circle (.2cm);
		\draw[<-,dotted, yshift=-.2cm] (\x,\x+1) -- ++(0, -.6); %^
		\draw[<-, dotted, yshift=-.2cm] (\x,\x) -- ++(0, -.6); %v
		\draw[<-, dotted, xshift=-.2cm] (\x,\x) -- ++(-.6, 0); %<
		\draw[<-, dotted, xshift=-.2cm] (\x+1,\x) -- ++(-.6, 0); %>
	}
}

\newcommand{\drawLabeledSquareBIS}[1]{
	\drawSquare{#1}
	\foreach \x in {1,...,#1}{
		\foreach \y in {1,...,#1}{
			\draw[opacity=0] (\x-1, \y-1) -- node[opacity=1, xshift=6pt] {\tiny $a_{\scalebox{.7}{2}}$} (\x-1, \y);
			\draw[opacity=0] (\x-1, \y-1) -- node[opacity=1, yshift=3pt] {\tiny$a_{\scalebox{.7}{1}}$} (\x, \y-1);
		}
	}
	\foreach \i in {1,...,#1} { 
		\draw[opacity=0] (\i-1, 0) -- node[opacity=1, xshift=6pt] {\tiny$a_{\scalebox{.7}{2}}$} (\i-1, -1);
		\draw[opacity=0] (0, \i-1) -- node[opacity=1, yshift=3pt] {\tiny $a_{\scalebox{.7}{2}}$} (-1, \i-1);
	}
}

\newcommand{\drawWhiteLine}[2]{
	\foreach
	[evaluate=\i as \x using {\i-1}]
	[evaluate=\i as \y using {mod(#1*\x,#2)}] 
	[evaluate=\i as \n using {mod(#1*120,360)}] 
	[evaluate=\i as \m using {mod(#1*120+30,360)}]
	\i in {1,...,#2} {
		\node[fill=white, draw=black,circle, minimum size=.4cm] (p) at (\x, \y) {};
	}
}

\newcommand{\drawLine}[2]{
	\foreach
	[evaluate=\i as \x using {\i-1}]
	[evaluate=\i as \y using {mod(#1*\x,#2)}] 
	[evaluate=\i as \n using {-1*#1*120-15}] 
	[evaluate=\i as \m using {-1*#1*120+15}]
	\i in {1,...,#2} {
		\node[circle, minimum size=.4cm] (p) at (\x, \y) {};
		\filldraw (p.center) --  (p.\n) arc (\n:\m:.2cm) -- (p.\m) -- cycle;
	}
}

\newcommand{\drawHard}{\showpictures{
		\begin{tikzpicture}
			\foreach \j in {0,1,2} {
				\foreach \i in {0, 4.5}
				\node[anchor=east] at (-1,\i+\j) {$q_2 {=} \j$};
				\foreach \i in {0, 4.5, 9, 13.5, 18}
				\node at (\i+\j,7.8) {$q_1 {=} \j$};
			}
			\node[anchor=west] at (21.5,5.5) {$q_4 {=} 1$};
			\node[anchor=west] at (21.5,1) {$q_4 {=} 0$};
			\node at (1,-2) {$q_3=0$};
			\node at (5.5,-2) {$q_3{=}1\quad C_1 {=} \{1\}$};
			\node at (10,-2) {$q_3{=}2\quad C_2 {=} \{1, 2\}$};
			\node at (14.5,-2) {$q_3{=}3\quad C_3 {=} \{2\}$};
			\node at (19,-2) {$q_3{=}4$};
			\draw [yshift=.2cm, decorate, decoration={brace,amplitude=10pt}]
			(21,7.5)++(0,-.4cm) -- (21,3.5);
			\draw [yshift=.2cm, decorate, decoration={brace,amplitude=10pt}]
			(21,3)++(0,-.4cm) -- (21,-1);
			
			\foreach
			[evaluate=\i as \j using {\i+4}]
			\i in {-1, 3.5, 8, 12.5, 17} {
				\draw [xshift=.2cm, decorate, decoration={brace,amplitude=10pt}]
				(\j,-1)++(-.4cm,0) -- (\i,-1);
			}
			
			\begin{scope}[xshift=0cm]
				\drawSquare{3}
			\end{scope}
			\begin{scope}[xshift=4.5cm]
				\drawSquare{3}
				\drawWhiteLine{1}{3}
				\drawLine{1}{3}
			\end{scope}
			\begin{scope}[xshift=9cm]
				\drawSquare{3}
				\drawWhiteLine{1}{3}
				\drawWhiteLine{2}{3}
				\drawLine{1}{3}
				\drawLine{2}{3}
			\end{scope}
			\begin{scope}[xshift=13.5cm]
				\drawSquare{3}
				\drawWhiteLine{2}{3}
				\drawLine{2}{3}
			\end{scope}
			\begin{scope}[xshift=18cm]
				\drawSquare{3}
			\end{scope}
			
			\begin{scope}[xshift=0cm,yshift=4.5cm]
				\drawSquare{3}
				\fill[black] (0,0) circle (.2cm);
			\end{scope}
			\begin{scope}[xshift=4.5cm,yshift=4.5cm]
				\drawSquare{3}
				\drawWhiteLine{1}{3}
				\drawLine{1}{3}
				\fill[black] (0,0) circle (.2cm);
			\end{scope}
			\begin{scope}[xshift=9cm,yshift=4.5cm]
				\drawSquare{3}
				\drawWhiteLine{1}{3}
				\drawWhiteLine{2}{3}
				\drawLine{1}{3}
				\drawLine{2}{3}
				\fill[black] (0,0) circle (.2cm);
			\end{scope}
			\begin{scope}[xshift=13.5cm,yshift=4.5cm]
				\drawSquare{3}
				\drawWhiteLine{2}{3}
				\drawLine{2}{3}
				\fill[black] (0,0) circle (.2cm);
			\end{scope}
			\begin{scope}[xshift=18cm,yshift=4.5cm]
				\drawSquare{3}
			\end{scope}
		\end{tikzpicture}
}}

\newcommand{\drawRequests}{\showpictures{
		\begin{tikzpicture}
			\providecommand\X{}
			\providecommand\Y{}
			\providecommand\R{}
			\providecommand\angle{}
			
			\renewcommand{\R}{2cm}

			\renewcommand{\X}{0}
			\renewcommand{\Y}{0}
			\node[threestate, fill=black] (A00) at ($(\X,\Y)$) {};
			\node[threestate] (A10) at ($(1*\R,0) + (\X,\Y)$) {};
			\node[threestate] (A01) at ($(0,-1*\R) + (\X,\Y)$) {};
			\node[threestate] (A11) at ($(1*\R,-1*\R) + (\X,\Y)$) {};
			
			\node[format] at (A00) {\color{white} 0,0};
			\node[format] at (A10) {1,0};
			\node[format] at (A01) {0,1};
			\node[format] at (A11) {1,1};
			
			\node at ($(\X - 1.5,\Y-0.5*\R)$) {$\A:$};
			\path[transition] ($(A00)+(-0.8,-0.8)$) edge (A00);
			
			\path[transition] (A00) edge [loop left] node {$g_1,g_2,i$} (A00);
			\path[transition] (A10) edge [loop right] node {$r_1,g_2,i$} (A10);
			\path[transition] (A01) edge [loop left] node {$r_2,g_1,i$} (A01);
			\path[transition] (A11) edge [loop right] node {$r_1,r_2$} (A11);

			\renewcommand{\angle}{10}
			
			\path[transition] (A00) edge [bend left=\angle] node[above] {$r_1$} (A10);
			\path[transition] (A10) edge [bend left=\angle] node[below] {$g_1$} (A00);
			
			\path[transition] (A01) edge [bend left=\angle] node[above] {$r_1$} (A11);
			\path[transition] (A11) edge [bend left=\angle] node[below] {$g_1$} (A01);
			
			\path[transition] (A00) edge [bend left=\angle] node[right] {$r_2$} (A01);
			\path[transition] (A01) edge [bend left=\angle] node[left] {$g_2$} (A00);
			
			\path[transition] (A10) edge [bend left=\angle] node[right] {$r_2$} (A11);
			\path[transition] (A11) edge [bend left=\angle] node[left] {$g_2$} (A10);

			\renewcommand{\X}{8}
			\node[threestate,fill=black] (B00) at ($(\X,\Y)$) {};
			\node[threestate] (B10) at ($(1*\R,0) + (\X,\Y)$) {};
			\node[threestate,fill=black] (B01) at ($(0,-1*\R) + (\X,\Y)$) {};
			\node[threestate] (B11) at ($(1*\R,-1*\R) + (\X,\Y)$) {};
			
			\node[format] at (B00) {\color{white} 0,x};
			\node[format] at (B10) {1,x};
			\node[format] at (B01) {\color{white} x,0};
			\node[format] at (B11) {x,1};
			
			\node at ($(\X - 2.5,\Y)$) {$\A_1:$};
			\node at ($(\X - 2.5,\Y-\R)$) {$\A_2:$};
			\path[transition] ($(B00)+(0,-0.7)$) edge [] (B00);
			\path[transition] ($(B01)+(0,0.7)$) edge [] (B01);
			
			\path[transition] (B00) edge [loop left] node {} (B00);
			\node at ($(\X-1.5,\Y+0.2)$) {$r_2,i$};
			\node at ($(\X-1.5,\Y-0.2)$) {$g_1,g_2$};
			\path[transition] (B10) edge [loop right] node {} (B10);
			\node at ($(1*\R,0) + (\X+1.5,\Y+0.2)$) {$g_2,i$};
			\node at ($(1*\R,0) + (\X+1.5,\Y-0.2)$) {$r_1,r_2$};
			\path[transition] (B01) edge [loop left] node {} (B01);
			\node at ($(0,-1*\R) + (\X-1.5,\Y+0.2)$) {$r_1,i$};
			\node at ($(0,-1*\R) + (\X-1.5,\Y-0.2)$) {$g_1,g_2$};
			\path[transition] (B11) edge [loop right] node {} (B11);
			\node at ($(1*\R,-1*\R) + (\X+1.5,\Y+0.2)$) {$g_1,i$};
			\node at ($(1*\R,-1*\R) + (\X+1.5,\Y-0.2)$) {$r_1,r_2$};

			\renewcommand{\angle}{10}
			
			\path[transition] (B00) edge [bend left=\angle] node[above] {$r_1$} (B10);
			\path[transition] (B10) edge [bend left=\angle] node[below] {$g_1$} (B00);

			\path[transition] (B01) edge [bend left=\angle] node[above] {$r_2$} (B11);
			\path[transition] (B11) edge [bend left=\angle] node[below] {$g_2$} (B01);

		\end{tikzpicture}
}}

\newcommand{\drawOrbits}{\showpictures{
		\begin{tikzpicture}
			
			\tikzset{every loop/.style={min distance=3ex, looseness=2}}
			
			\providecommand\X{}
			\providecommand\Y{}
			\providecommand\R{}
			
			\renewcommand{\R}{1.3cm}

			\renewcommand{\X}{-1.25}
			\renewcommand{\Y}{0}
			\node[state,fill=black] (A1) at ($(-0.75,0.4*\R) + (\X,\Y)$) {};
			\node[state] (A2) at ($(0,0.85*\R) + (\X,\Y)$) {};
			\node[state] (A3) at ($(0.75,0.4*\R) + (\X,\Y)$) {};
			\node[state] (A4) at ($(0.75,-0.4*\R) + (\X,\Y)$) {};
			\node[state] (A5) at ($(0,-0.85*\R) + (\X,\Y)$) {};
			\node[state] (A6) at ($(-0.75,-0.4*\R) + (\X,\Y)$) {};
			
			\node[] at (A1) {\color{white} 1};
			\node[] at (A2) {2};
			\node[] at (A3) {3};
			\node[] at (A4) {4};
			\node[] at (A5) {5};
			\node[] at (A6) {6};
			
			\node at ($(\X-2.05,\Y)$) {$\A:$};
			
			\path[transition] ($(A1)+(-0.75,0)$) edge [] (A1);
			
			\path[transition] (A5) edge [loop right,looseness=2,min distance=3ex, in=0, out=330] node[right] {\scriptsize$b$} (A5);
			\path[transition] (A2) edge [loop left,looseness=2,min distance=3ex, in=150, out=180] node[left] {\scriptsize$b$} (A2);
			
			\path[transition] (A2) edge node[above left,yshift=-2pt] {\scriptsize$a$} (A1);
			\path[transition] (A1) edge node[below] {\scriptsize$a$} (A3);
			\path[transition] (A3) edge node[above right,yshift=-2pt] {\scriptsize$a$} (A2);
			
			\path[transition] (A6) edge node[below left,yshift=1pt] {\scriptsize$a$} (A5);
			\path[transition] (A5) edge node[below right,yshift=1pt] {\scriptsize$a$} (A4);
			\path[transition] (A4) edge node[above] {\scriptsize$a$} (A6);

			\path[bitransition] (A6) edge node[left] {\scriptsize$b$} (A1);
			\path[bitransition] (A4) edge node[right] {\scriptsize$b$} (A3);

			\renewcommand{\X}{0.}
			\node[] (A1) at ($(-0.75,0.85*\R) + (\X,\Y)$) {};
			\node[] (A2) at ($(0,0.85*\R) + (\X,\Y)$) {};
			\node[] (A3) at ($(0.75,0.85*\R) + (\X,\Y)$) {};
			\node[] (A4) at ($(0.75,-0.85*\R) + (\X,\Y)$) {};
			\node[] (A5) at ($(0,-0.85*\R) + (\X,\Y)$) {};
			\node[] (A6) at ($(-0.75,-0.85*\R) + (\X,\Y)$) {};
			
			\renewcommand{\X}{7.5}
			\renewcommand{\Y}{0.}
			\node[threestate] (B234) at ($(A3) + (\X,\Y)$) {};
			\node[threestate] (B456) at ($(A5) + (\X,\Y)$) {};
			\node[threestate,fill=black] (B126) at ($(A1) + (\X,\Y)$) {};
			\node[threestate,fill=black] (B135) at ($(\X,\Y)$) {};

			\node[] at (B234) {234};
			\node[] at (B456) {456};
			\node[] at (B126) {\color{white}126};
			\node[] at (B135) {\color{white}135};
			
			\path[transition] (B456) edge [loop right] node {\scriptsize$a$} (B456);
			\path[transition] (B234) edge node[above] {\scriptsize$a$} (B126);
			\path[transition] (B126) edge node[below left, yshift=5pt] {\scriptsize$a$} (B135);
			\path[transition] (B135) edge node[below right, yshift=5pt] {\scriptsize$a$} (B234);
			\path[bitransition] (B135) edge node[left] {\scriptsize$b$} (B456);
			\path[transition] (B234) edge [loop right] node {\scriptsize$b$} (B234);
			\path[transition] (B126) edge [loop left] node {\scriptsize$b$} (B126);

			\renewcommand{\X}{5.}
			\renewcommand{\Y}{-0.}
			\node[threestate,fill=black] (B123) at ($(A2) + (\X,\Y)$) {};
			\node[threestate,fill=black] (B156) at ($(A6) + (\X,\Y)$) {};
			\node[threestate] (B345) at ($(A4) + (\X,\Y)$) {};
			\node[threestate] (B246) at ($(\X,\Y)$) {};

			\node[] at (B123) {\color{white} 123};
			\node[] at (B156) {\color{white} 156};
			\node[] at (B345) {345};
			\node[] at (B246) {246};
			
			\path[transition] (B123) edge [loop left] node {\scriptsize$a$} (B123);
			\path[transition] (B156) edge node[above] {\scriptsize$a$} (B345);
			\path[transition] (B345) edge node[above right, yshift=-5pt] {\scriptsize$a$} (B246);
			\path[transition] (B246) edge node[above left, yshift=-5pt] {\scriptsize$a$} (B156);
			\path[bitransition] (B123) edge node[left] {\scriptsize$b$} (B246);
			\path[transition] (B156) edge [loop left] node {\scriptsize$b$} (B156);
			\path[transition] (B345) edge [loop right] node {\scriptsize$b$} (B345);

			\renewcommand{\X}{2.5}
			\renewcommand{\Y}{+0.}
			\node[threestate,fill=black] (B14) at ($(A1) + (\X,\Y)$) {};
			\node[threestate] (B36) at ($(A3) + (\X,\Y)$) {};
			\node[threestate] (B25) at ($(A5) + (\X,\Y)$) {};
			
			\node[] at (B14) {\color{white} 14};
			\node[] at (B36) {36};
			\node[] at (B25) {25};
			
			\path[transition] (B25) edge [loop left] node {\scriptsize$b$} (B25);
			\path[transition] (B25) edge node[left] {\scriptsize$a$} (B14);
			\path[transition] (B14) edge [bend left] node[above] {\scriptsize$a$} (B36);
			\path[transition] (B36) edge node[right] {\scriptsize$a$} (B25);
			\path[bitransition] (B36) edge [bend left] node[below] {\scriptsize$b$} (B14);

			\renewcommand{\X}{10}
			\renewcommand{\Y}{-0.}
			\node[threestate] (B2356) at ($(A4) + (\X,\Y)$) {};
			\node[threestate,fill=black] (B1245) at ($(A6) + (\X,\Y)$) {};
			\node[threestate,fill=black] (B1346) at ($(A2) + (\X,\Y)$) {};
			
			\node[] at (B2356) {2356};
			\node[] at (B1245) {\color{white} 1245};
			\node[] at (B1346) {\color{white} 1346};

			\path[transition] (B1346) edge [loop right] node {\scriptsize$b$} (B1346);
			\path[transition] (B1346) edge node[right] {\scriptsize$a$} (B2356);
			\path[transition] (B2356) edge [bend left] node[below] {\scriptsize$a$} (B1245);
			\path[transition] (B1245) edge node[left] {\scriptsize$a$} (B1346);
			\path[bitransition] (B1245) edge [bend left] node[above] {\scriptsize$b$} (B2356);
			
		\end{tikzpicture}
}}

\newcommand{\drawDecomposition}{\showpictures{
		\begin{tikzpicture}
			\begin{scope}[xshift=-3cm]
				\node[] (N) at (-1.25,1.8) {$\A_{5}^2:$};
				\scalebox{.9}{\drawLabeledSquareBIS{5}}
				\scalebox{.9}{\drawWhiteLine{1}{5}}
				\scalebox{.9}{\drawWhiteLine{2}{5}}
				\scalebox{.9}{\drawWhiteLine{3}{5}}
				\scalebox{.9}{\drawWhiteLine{4}{5}}
			\end{scope}
			\begin{scope}[xshift=3cm]
				\foreach \y in {1,...,4}{
					\node[] (N\y) at (0.25,5.2-1.3*\y) {$\A^2_{5.\y}:$};
					\node[fill=white, draw=black,circle, minimum size=.2cm](A1\y) at (1,5.2-1.3*\y) {};
					\foreach \x/\z in {2/1,3/2,4/3,5/4}{
						\node[fill, circle, minimum size=.2cm](A\x\y) at (\x,5.2-1.3*\y) {};
						\ifthenelse{\y = 1}{
							\draw[trans] (A\z\y) edge[bend left] node[lowlab] {\tiny $a_{\scalebox{.7}1}$}(A\x\y); %>
						}
						{
							\ifthenelse{\y = 4}{
								\draw[trans] (A\z\y) -- node[lowlab] {\tiny $a_{\scalebox{.7}1},a_{\scalebox{.7}2}$}(A\x\y); %>
							}
							{
								\draw[trans] (A\z\y) -- node[lowlab] {\tiny $a_{\scalebox{.7}1}$}(A\x\y); %>
							}
						}
					}
					\ifthenelse{\y = 4}{
						\draw[trans] (A5\y) edge [bend left=30] node[lowlab] {\tiny $a_{\scalebox{.7}1},a_{\scalebox{.7}2}$} (A1\y);
					}
					{
						\draw[trans] (A5\y) edge [bend left=30] node[lowlab] {\tiny $a_{\scalebox{.7}1}$} (A1\y);
					}
				}
				
				\draw[trans] (A11) edge [bend left=30] node[lowlab] {\tiny $a_{\scalebox{.7}2}$} (A51);
				\draw[trans] (A21) edge [bend left=30] node[lowlab] {\tiny $a_{\scalebox{.7}2}$} (A11);
				\draw[trans] (A31) edge [bend left=30] node[lowlab] {\tiny $a_{\scalebox{.7}2}$} (A21);
				\draw[trans] (A41) edge [bend left=30] node[lowlab] {\tiny $a_{\scalebox{.7}2}$} (A31);
				\draw[trans] (A51) edge [bend left=30] node[lowlab] {\tiny $a_{\scalebox{.7}2}$} (A41);
				
				\draw[trans] (A12) edge [bend left=30] node[lowlab] {\tiny $a_{\scalebox{.7}2}$} (A42);
				\draw[trans] (A22) edge [bend left=30] node[lowlab] {\tiny $a_{\scalebox{.7}2}$} (A52);
				\draw[trans] (A32) edge [bend left=30] node[lowlab] {\tiny $a_{\scalebox{.7}2}$} (A12);
				\draw[trans] (A42) edge [bend left=30] node[lowlab] {\tiny $a_{\scalebox{.7}2}$} (A22);
				\draw[trans] (A52) edge [bend left=30] node[lowlab] {\tiny $a_{\scalebox{.7}2}$} (A32);
				
				\draw[trans] (A13) edge [bend left=30] node[lowlab] {\tiny $a_{\scalebox{.7}2}$} (A33);
				\draw[trans] (A23) edge [bend left=30] node[lowlab] {\tiny $a_{\scalebox{.7}2}$} (A43);
				\draw[trans] (A33) edge [bend left=30] node[lowlab] {\tiny $a_{\scalebox{.7}2}$} (A53);
				\draw[trans] (A43) edge [bend left=30] node[lowlab] {\tiny $a_{\scalebox{.7}2}$} (A13);
				\draw[trans] (A53) edge [bend left=30] node[lowlab] {\tiny $a_{\scalebox{.7}2}$} (A23);
				
			\end{scope}
		\end{tikzpicture}
}}

\newcommand{\myparagraph}[1]{\vskip0.7\baselineskip\noindent\textcolor{lipicsGray}{\sffamily\bfseries#1\enskip}}
\newcommand{\boldtitle}[1]{\noindent\textcolor{lipicsGray}{\sffamily\bfseries#1\enskip}}

\global\long\def\zug#1{\langle#1\rangle}
\global\long\def\st{\mid}

\global\long\def\width{{\it width}}
\newcommand{\HIT}{\textsf{HIT}\xspace}
\newcommand{\shortish}{concise\xspace}

\newcommand{\A}{\mathcal{A}}
\newcommand{\B}{\mathcal{B}}
\newcommand{\C}{\mathcal{C}}
\newcommand{\N}{\mathbb{N}}

\newcommand{\DFA}{\textsf{DFA}\xspace}
\newcommand{\DFAs}{\textsf{DFAs}\xspace}

\newcommand{\NP}{\textsf{NP}\xspace}

\newcommand{\FPT}{\textsf{FPT}\xspace}
\newcommand{\PSPACE}{\textsf{PSPACE}\xspace}
\newcommand{\NPSPACE}{\textsf{NPSPACE}\xspace}
\newcommand{\boundDecomp}{\textsc{Bound-Decomp}\xspace}
\newcommand{\Decomp}{\textsc{Decomp}\xspace}
\newcommand{\NLOGSPACE}{\textsf{NL}\xspace}
\newcommand{\LOGSPACE}{\textsf{LOGSPACE}\xspace}
\newcommand{\PTIME}{\textsf{PTIME}\xspace}

\newcommand{\EXPSPACE}{\textsf{EXPSPACE}\xspace}

\newcommand{\affects}{:=}
\newcommand{\testeq}{\ensuremath{\mathrel{\mbox{\small ?}{=}}}}
\newcommand{\appendixProof}{appendix}
\newcommand{\p}{\ensuremath{\mu}}
\newcommand{\q}{\ensuremath{\tau}}
\newcommand{\commentApx}[2]{\IfAppendix{#2}{#1}}

\makeatletter 
\xpatchcmd{\thmt@restatable}% Edit \thmt@restatable
{\csname #2\@xa\endcsname\ifx\@nx#1\@nx\else[{#1}]\fi}% Replace this code
{\IfAppendix{\csname #2\@xa\endcsname}{\csname #2\@xa\endcsname\ifx\@nx#1\@nx\else[{#1}]\fi}}% with this code
{}{} % execute code for success/failure instances
\makeatother

\usepackage{todonotes}

\begin{document}
	\maketitle
	\begin{abstract}
		A deterministic finite automaton (\DFA)
		$\A$ is composite if its language $L(\A)$
		can be decomposed into an intersection
		$\bigcap_{i=1}^k L(\A_i)$
		of languages of smaller \DFAs. 
		Otherwise, $\A$ is prime.
		This notion of primality was introduced by
		Kupferman and Mosheiff in 2013, and while they proved
		that we can decide whether
		a \DFA is composite, the precise
		complexity of this problem is still open,
		with a doubly-exponential gap
		between the upper and lower bounds.
		In this work, we focus on permutation \DFAs, i.e., those for which the transition monoid is a group.
		We provide an \NP algorithm
		to decide whether a permutation 
		\DFA is composite,
		and show that the difficulty of this
		problem comes from the number of non-accepting states of the instance:
		we give a fixed-parameter tractable 
		algorithm
		with
		the number of rejecting states as the parameter.
		Moreover, we investigate the class of commutative permutation \DFAs.
		Their structural properties
		allow us to decide compositionality in \NLOGSPACE,
		and even in \LOGSPACE if the alphabet size is fixed.
		Despite this low complexity,
		we show that complex behaviors still
		arise in this class:
		we provide a family of composite \DFAs each requiring polynomially many factors with respect to its size.
		%we define a family $(\A_n^m)_{m,n \in \N}$ of composite \DFAs of size $n^m$ that require $\mathcal{O}(n^{m-1})$ \DFAs in their decomposition.
		We also consider
		the variant of the problem that asks
		whether a \DFA is \emph{$k$-factor composite},
		that is,
		decomposable into $k$ smaller \DFAs,
		for some given integer $k \in \N$.
		We show that, for commutative permutation \DFAs,
		restricting the number of factors makes the decision computationally harder,
		and yields a problem with tight bounds: it is \NP-complete.
		Finally, we show that in general, this problem is in \PSPACE,
		and it is in \LOGSPACE for \DFAs with a singleton alphabet.
	\end{abstract}

	\section{Introduction}
	Compositionality is a fundamental notion in
	numerous fields of computer science~\cite{dRLP98}.
	This principle can be summarised as follows:
	Every system should be designed by composing
	simple parts
	such that the meaning of the system
	can be deduced from the meaning of its parts,
	and how they are combined.
	For instance, this is a crucial aspect of modern
	software engineering:
	a program
	split into simple modules
	will be quicker to compile
	and easier to maintain.
	The use of compositionality is also
	essential in theoretical computer science:
	it is used to avoid
	the \emph{state explosion} issues that
	usually happen when combining
	parallel processes together,
	and also to overcome the \emph{scalability}
	issues of problems with
	a high theoretical complexity.
	In this work, we study compositionality
	in the setting of formal languages:
	we show how to make languages simpler
	by decomposing them into \emph{intersections} of smaller languages. 
	This is motivated by the \emph{model-checking} problems.
	For instance, the \textsf{LTL} model-checking problem 
	asks, given a linear temporal logic formula $\varphi$
	and a finite state machine $M$,
	whether every execution of $M$ satisfies $\varphi$.
	This problem is decidable,
	but has a high theoretical complexity (\PSPACE)
	with respect to the size of $\varphi$~\cite{DBLP:books/daglib/0020348}.
	If $\varphi$ is too long,
	it cannot be checked efficiently.
	This is where compositionality comes into play:
	if we can decompose the specification language into an intersection of simple languages,
	that is, decompose $\varphi$
	into a conjunction $\varphi =
	\varphi_1 \wedge \varphi_2 \wedge \cdots \wedge \varphi_k$
	of small specifications,
	it is sufficient to check
	whether all the $\varphi_i$ are satisfied separately.

	\begin{figure}[]
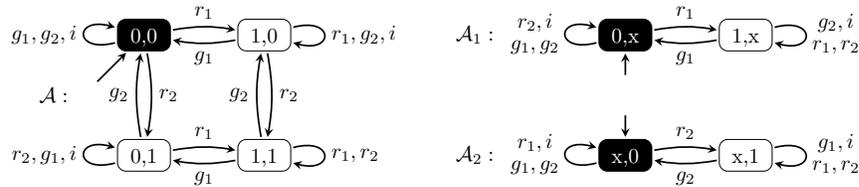

		\centering
		\scalebox{.8}{\drawRequests}
		\caption{\DFAs
			recognising specifications.
			Accepting states are drawn in black.
			The \DFAs $\A_1$ and $\A_2$
			check that every request of
			the first, resp.~second, client is eventually granted,
			$\A$ checks both.} \label{ex:request}
	\end{figure}

	Our aim is to develop
	the theoretical foundations
	of the compositionality principle for formal languages
	by investigating
	how to decompose into simpler parts
	one of the most basic model
	of abstract machines:
	deterministic finite automata (\DFAs).
	We say that a \DFA $\A$ is \emph{composite}
	if its language can be decomposed
	into the intersection of the languages
	of smaller $\DFAs$.
	More precisely, we say that $\A$ is
	$k$-\emph{factor composite}
	if there exist
	$k$ \DFAs $(\A_i)_{1 \leq i \leq k}$
	with less states than $\A$
	such that $L(\A) = \bigcap_{i=1}^k L(\A_i)$.
	We study the two following problems:
	
	\begin{center}
		\begin{minipage}[l]{.5\linewidth-.6cm}
			\textsc{\DFA} \Decomp \ \\
			Given: \DFA $\A$.\\
			Question: Is $\A$ composite?
		\end{minipage}
		\begin{minipage}[l]{.5\linewidth-.6cm}
			\textsc{\DFA} \boundDecomp\ \\
			Given: \DFA $\A$ and integer $k \in \mathbb{N}$.\\
			Question: Is $\A$ $k$-factor composite?
		\end{minipage}
	\end{center}

	%These problems are motivated by
	%the automata-theoretic approach to formal verification~\cite{DBLP:conf/ijcai/GiacomoV15,DBLP:conf/hvc/RozierV12,Var95c}:
	%\ismael{Say something about finite vs infinite words?}
	%\nicolas{\cite{DBLP:conf/hvc/RozierV12} speak about the inherent of determinism in safety specifications\\\cite{DBLP:conf/ijcai/GiacomoV15} is about LTL and LDL over finite words}
	%as stated before, the model-checking problem
	%is easier to solve when we can
	%decompose the specification 
	%into a \emph{conjunction} of smaller specifications.
	%For specifications defined by \DFAs,
	%determining the existence of such a decomposition amounts to solving the \textsc{Decomposition}
	%problem.
	\noindent
	The next example shows that
	decomposing \DFAs
	can result in substantially smaller machines.
	
	\myparagraph{Example}
	Consider Figure \ref{ex:request}.
	We simulate the interactions between
	a system and two clients
	by using finite words
	on the alphabet $\{r_1,r_2,g_1,g_2,i\}$:
	At each time step,
	the system either
	receives a request from a client ($r_1, r_2$),
	grants the open requests of a client ($g_1,g_2$),
	or stays idle ($i$).
	A basic property usually required
	is that every request is eventually granted.
	This specification is recognised
	by the \DFA $\A$, which keeps
	track in its state of the current open
	requests, and only accepts if
	none is open when the input ends.
	Alternatively, 
	this specification
	can be decomposed into
	the intersection
	of the languages defined by the \DFAs
	$\A_1$ and $\A_2$: each one
	checks that the requests
	of the corresponding client
	are eventually granted.
	While in this precise example
	both ways of defining the specification
	are comparable,
	the latter scales drastically better than the former
	when the number of clients increases:
	Suppose that there are now
	$n \in \mathbb{N}$ clients.
	In order to check that all the
	requests are granted with 
	a single \DFA, we need $2^n$ states
	to keep track of all possible
	combinations of open requests,
	which is impractical when $n$ gets too big.
	However, decomposing this specification into
	an intersection yields $n$ \DFAs of size two,
	one for each client.
	Note that, while in this specific example the decomposition is obvious, in general computing such a conjunctive form can be challenging: currently the best known algorithm needs exponential space.
	
	\myparagraph{\DFAs in hardware} Our considered problems are of great interest in hardware implementations of finite state machines~\cite{pedroni2013finite} where realizing large \DFAs poses a challenge~\cite{gould2007apparatus}. In~\cite{clarke1991language} the authors describe a state machine language for describing complex finite state hardware controllers, where the compiled state tables can automatically be input into a temporal logic model checker.
	If the control mechanism of the initial finite state machine can be split up into a conjunction of constraints, considering a decomposition instead could improve this work-flow substantially.
	Decomposing a complex \DFA $\A$ can lead to a smaller representation of the \DFA in total,
	as demonstrated in the previous example in Figure~\ref{ex:request},
	and on top of that the individual smaller \DFAs $\A_i$ in the decomposition $L(\A) = \bigcap_{i=1}^k L(\A_i)$
	can be placed independently on a circuit board, as they do not have to interact with each other and only need to read their common input from a global bus and signal acceptance as a flag to the bus. 
	This allows for a great flexibility in circuit designs, as huge \DFAs can be broken down into smaller blocks which fit into niches giving space for inflexible modules such as CPU cores.
	%The acceptance of the initial \DFA can then be realized as a conjunction of the acceptances of the factor \DFAs.
	
	\myparagraph{Reversible \DFAs}
	We focus our study on \emph{permutation} \DFAs, which are \DFAs whose transition monoids are groups:
	each letter induces a one-to-one map from the state set into itself.
	These \DFAs are also called \emph{reversible} \DFAs~\cite{DBLP:conf/mfcs/KuncO13,DBLP:conf/latin/Pin92}.
	Reversibility is stronger than determinism:
	this powerful property allows to deterministically navigate \emph{back and forth} between the steps of a computation.
	This is particularly relevant in the study of the physics of computation,
	since irreversibility causes energy dissipation~\cite{DBLP:journals/ibmrd/Landauer61}.
	Remark that in the setting of \DFAs, this power results in a loss of expressiveness:
	contrary to more powerful models (for instance Turing machines),
	reversible \DFAs are less expressive than general \DFAs.
	
	\myparagraph{Related work}
	The \DFA \Decomp problem
	was first introduced in 2013
	by Kupferman and Moscheiff~\cite{KM15}.
	They proved that it
	is decidable in \EXPSPACE,
	but left open the exact
	complexity:
	the best known lower bound is hardness for
	\NLOGSPACE.
	They gave more efficient
	algorithms for restricted
	domains:
	a \PSPACE algorithm
	for \emph{permutation} \DFAs,
	and a \PTIME algorithm
	for \emph{normal} permutation \DFAs,
	a class of \DFAs that contains all \emph{commutative} permutation \DFAs.
	Recently, the \Decomp
	problem was proved
	to be decidable in \LOGSPACE
	for \DFAs with a singleton alphabet~\cite{JKM20}.
	%The complexity of the \boundDecomp problem has not been studied directly,
	%but the trade-off between the number of factors and their sizes has been studied:
	%some \DFAs can either be decomposed in a large number of small factors,
	%or in a small number of large factors~\cite{netser}.
	%
	The trade-off between number and size of factors was studied in~\cite{netser}, where automata showing extreme behavior are presented, i.e., \DFAs that can either be decomposed into a large number of small factors,
	or a small number of large factors.

	\begin{figure}[]
		\centering
		\begin{tabular}{|r||c|c|}
			\hline
			&
			\makebox[3.8cm]{\Decomp}&
			\makebox[3.8cm]{\boundDecomp}
			\\\hline\hline
			\DFAs
			&\EXPSPACE~\cite{KM15}&{\bf \PSPACE}\\\hline
			Permutation \DFAs 
			&{\bf \NP/\FPT}&{\bf \PSPACE}\\\hline
			Commutative permutation \DFAs 
			&{\bf \NLOGSPACE}&{\bf \NP-complete}\\\hline
			Unary \DFAs 
			&\LOGSPACE~\cite{JKM20}&{\bf \LOGSPACE}\\\hline
		\end{tabular}
		\caption{Complexity of studied problems with containing classes, with our contribution in {\bf bold}.} \label{ex:table}
	\end{figure}
	
	\myparagraph{Contribution}
	We expand the domain of instances over which the
	\Decomp problem is tractable.
	We focus on permutation \DFAs, and we propose
	new techniques that improve the known complexities.
	All proofs omitted due to space restrictions can be found in the full version.
	Unless specified otherwise,
	the complexity of our algorithms do not depend on the size of the alphabet of the \DFA.
	Our results, summarised by Figure \ref{ex:table},
	are presented as follows.
	
	\boldtitle{Section \ref{sec:permutation}:}
	We give an \NP algorithm
	for permutation \DFAs,
	and we show that the complexity
	is directly linked to the
	number of non-accepting states.
	This allows us to obtain
	a fixed-parameter tractable
	algorithm with respect to the number
	of non-accepting states (Theorem~\ref{thm:permutation_automata_prime_NP}).
	Moreover, we prove that permutation
	\DFAs with a prime number of states
	cannot be decomposed (Theorem~\ref{theorem:prime_is_prime}).
	
	\boldtitle{Section \ref{sec:commutative}:}
	We consider \emph{commutative}
	permutation \DFAs,
	where the \Decomp problem
	was already known to be tractable,
	and we lower the complexity from \PTIME to
	\NLOGSPACE, and even \LOGSPACE if the size of the alphabet is fixed (Theorem \ref{thm:commutative_automata_prime_LS}).
	While it is easy to decide whether a 
	commutative permutation \DFA is composite,
	we show that
	rich and complex behaviours
	still appear in this class:
	there exist families of composite \DFAs that require
	polynomially many factors to get a decomposition.
	More precisely, we construct a family $(\A_n^m)_{m,n \in \N}$ of composite \DFAs such that $\A_n^m$ is a \DFA of size $n^m$ 
	that is $(n-1)^{m-1}$-factor composite but not $(n-1)^{m-1}-1$-factor composite (Theorem \ref{theorem:poly_family}).
	Note that, prior to this result,
	only families of composite \DFAs with sublogarithmic
	width were known~\cite{JKM20}.
	
	\boldtitle{Section \ref{sec:bounded_decomposition}:}
	Finally,
	we study the \boundDecomp problem.
	High widths are undesirable
	for practical purposes:
	dealing with a huge number of small \DFAs
	might end up being more complex
	than dealing with a single \DFA of moderate size.
	The \boundDecomp problem
	copes with this issue
	by limiting the number of factors
	allowed in the decompositions.
	We show that this flexibility comes
	at a cost:
	somewhat surprisingly, this
	problem is \NP-complete
	for commutative permutation \DFAs (Theorem \ref{thm:bounded_dec_NP}),
	a setting where the
	\Decomp problem is easy.
	We also show that this problem is in \PSPACE for the general setting (Theorem \ref{thm:bounded_decomposition}), and in \LOGSPACE for unary \DFAs i.e.\ with a singleton alphabet (Theorem \ref{thm:bounded_decomposition_unary}).
	\section{Definitions}
	We denote by $\N$ the set of non-negative integers
	$\{0,1,2,\ldots\}$.
	%For an integer $n\in \mathbb{N}$ we define $[n]=\{1, 2, \dots, n\}$.\notaN{Check where "$[n]$" is not used}
	For a word $w=w_1w_2 \dots w_n$ with $w_i \in \Sigma$ for $1 \leq i \leq n$, we denote with $w^R=w_n\dots w_2w_1$ the \emph{reverse} of $w$.
	Moreover, for every $\sigma \in \Sigma$,
	we denote by $\#_\sigma(w)$ the number of times the letter $\sigma$ appears in $w$. 
	A natural number $n > 1$
	is called \emph{composite}
	if it is the product of two smaller numbers,
	otherwise we say that $n$ is \emph{prime}.
	Two integers $m,n \in \mathbb{N}$
	are called \emph{co-prime}
	if their greatest common divisor is $1$.
	We will use the following
	well known results~\cite{hardy1929,doi:10.4169/amer.math.monthly.120.07.650}:
	
	\boldtitle{Bertrand's Postulate:}
	For all $n>3$
	there is a prime number $p$
	satisfying $n<p<2n-2$.
	
	\boldtitle{B\'ezout's Identity:}
	For every pair of integers $m,n \in \mathbb{N}$,
	the set $\{\lambda m - \mu n \st \lambda,\mu \in \mathbb{N}\}$ contains exactly the multiples
	of the greatest common divisor of $m$ and $n$.

	\myparagraph{Deterministic finite automata}
	A \emph{deterministic finite automaton} (\DFA hereafter) is a $5$-tuple
	$\A=\zug{\Sigma,Q,q_I,\delta,F}$, where $Q$ is a finite set of states,
	$\Sigma$ is a finite non-empty alphabet, $\delta \colon Q\times\Sigma\to Q$
	is a transition function, $q_I\in Q$ is the initial state, and $F\subseteq Q$
	is a set of accepting states.
	The states in $Q \setminus F$ are called \emph{rejecting} states.
	%For each state $q\in Q$,
	%we use $\A^{q}$ to denote the \DFA $\A$
	%with $q$ as the initial state.
	%That is, $\A^{q}=\zug{\Sigma,Q,q,\delta,F}$. 
	We extend $\delta$ to words in the expected way,
	thus $\delta \colon Q\times\Sigma^{*}\to Q$ is defined recursively by
	$\delta(q,\varepsilon)=q$ and
	$\delta(q,w_{1}w_{2}\cdots w_{n})=\delta(\delta(q,w_{1}w_{2}\cdots w_{n-1}),w_{n})$.
	% We sometimes omit the initial state $q_I$ as a parameter of $\delta$ and 
	% write $\delta(w)$ instead of $\delta(q_I,w)$ in order to refer to the state that $\A$ visits after reading $w$.
	% The \DFA $\A$ naturally induces an equivalence relation $\sim_\A$ over
	% the set of words $\Sigma^*$ defined by
	% $u \sim_A v$ iff $\delta(u) = \delta(v)$.
	%
	The \emph{run} of $\A$ on a word $w=w_1\ldots w_n$
	is the sequence of states $s_{0}, s_{1}, \dots, s_{n}$
	such that $s_{0}=q_{I}$ and for each $1\leq i\leq n$ it holds that
	$\delta(s_{i-1},w_{i})=s_{i}$.
	Note that $s_n=\delta(q_I,w)$.
	The \DFA $\A$ \emph{accepts} $w$ iff $\delta(q_I,w)\in F$.
	Otherwise, $\A$ \emph{rejects}
	$w$.
	The set of words accepted by $\A$ is denoted $L(\A)$
	and is called the \emph{language of $\A$}. 
	%We say that $\A$ \emph{recognizes} $L(\A)$. 
	A language accepted by some \DFA is called
	a \emph{regular language}.
	% Two \DFAs $\A$ and $\B$ are \emph{equivalent} if $L(\A)=L(\B)$.
	
	We refer to the size of a \DFA $\A$, denoted $|\A|$,
	as the number of states in $\A$.
	A \DFA $\A$ is \emph{minimal}
	if every \DFA $\B$ such that $L(\B) = L(\A)$ satisfies $|\B| \geq |\A|$.
	%Every regular
	%language $L$ has a single (up to isomorphism)
	%minimal \DFA $\A$ such that
	%$L(\A)=L$. The index of $L$, denoted $\ind(L)$,
	%is the size of the minimal \DFA recognizing $L$. 

	\myparagraph{Composite \DFAs}
	We call a \DFA $\A$ \emph{composite} if there exists a family $(\B_i)_{1 \leq i \leq k}$
	of \DFAs with $|\B_i| < |\A|$ for all $1 \leq i \leq k$ such that $L(\A) = \bigcap_{1 \leq i \leq k}L(\B_i)$ and 
	call the family $(\B_i)_{1 \leq i \leq k}$ a \emph{decomposition} of $\A$. 
	Note that, all $\B_i$ in the decomposition satisfy $|\B_i| <|\A|$ and $L(\A)\subseteq L(\B_i)$.
	Such \DFAs are called \emph{factors} of $\A$, and $(\B_i)_{1 \leq i \leq k}$
	is also called a \emph{$k$-factor decomposition} of $\A$.
	The \emph{width} of $\A$
	%denoted as $\width(\A)$
	is the smallest $k$ for which there is a $k$-factor decomposition of $\A$,
	and we say that $\A$ is $k$-\emph{factor composite} iff $\width(\A) \leq k$.
	%If for all $1 \leq i \leq n, |\B_i| \leq k$, we call $\B_1, \B_2, \dots, \B_n$ a \emph{$k$-decomposition} of $\A$.\notaN{Check if "$k$-decomposition" is used}
	%Further, we define the \emph{depth} af $\A$, denoted as $\depth(\A)$, as the minimal $k$ for which there is a $k$-decomposition of $\A$.
	We call a \DFA $\A$ \emph{prime} if it is not composite.
	We call a \DFA $\A$ \emph{trim} if 
	all of its states are accessible from the initial state.
	As every non-trim \DFA $\A$ is composite,
	we assume all given \DFAs to be trim in the following. 
	
	We call a \DFA a \emph{permutation \DFA} if for each letter $\sigma \in \Sigma$,
	the function mapping each state $q$ to the state $\delta(q,\sigma)$ is a bijection.
	For permutation \DFAs the transition monoid is a group.
	Further, we call a \DFA $\A=\zug{\Sigma,Q,q_I,\delta,F}$ a  \emph{commutative \DFA}
	if 
	$\delta(q,uv) = \delta(q,vu)$
	for every state $q$ 
	and every pair of words $u,v \in \Sigma^*$.
	%We call a permutation \DFA \emph{normal} iff  for every $u \in F_\Sigma$ such that $\delta_u$ has a fixed point, it holds that $\delta_u$ is the identity function on $Q$, where $F_\Sigma$ is the group generated by $\Sigma \cup \{\sigma^{-1} \st \sigma \in \Sigma\}$. \todo[color=magenta]{Todo finish this definition, what is the slimmest one?}
	%Note that every commutative permutation \DFA is a normal permutation \DFA.
	In the next sections we discuss the problem of being composite for the classes of permutation \DFA,
	and commutative permutation \DFAs. 
	
	\section{Decompositions of Permutation \DFAs}
	\label{sec:permutation}
	In this section, we study permutation \DFAs.
	Our main contribution is an algorithm
	for the \Decomp problem
	that is \FPT with respect to the number of rejecting states:
	
	\begin{theorem}\label{thm:permutation_automata_prime_NP}
		The \Decomp problem for 
		permutation \DFAs is in \NP.
		It is in \FPT with parameter $k$, being the number of rejecting states of \DFA $\A$, solvable in time $\mathcal{O}(2^kk^2 \cdot|\A|)$.
		%	The primality problem is \NP for permutation \DFA.
		%	\orange{It is \textsc{PTime} when the number of rejecting states is fixed.}\notaN{\towrite FPT with $Q \setminus F$}
	\end{theorem}
	
	\noindent
	We prove Theorem~\ref{thm:permutation_automata_prime_NP}
	by introducing the notion of \emph{orbit-\DFAs}:
	an orbit-\DFA $\A^U$ of a \DFA $\A$
	is the \DFA
	obtained by fixing a set of states $U$ of $\A$ as the initial state,
	and letting the transition function
	of $\A$ act over it (thus the states of $\A^U$
	are subsets of the state space of $\A$).
	We prove three key results:
	\begin{itemize}
		\item 
		A permutation \DFA is composite
		if and only if it can be decomposed into
		its orbit-\DFAs
		(Corollary~\ref{cor:permutation_orbit_decomposition});
		\item 
		A permutation \DFA $\A$ can be decomposed into
		its orbit-\DFAs
		if and only if for each of its rejecting 
		states $q$,
		there exists an orbit-\DFA $\A^U$
		smaller than $\A$
		that \emph{covers} $q$, that is,
		one of the states of $\A^U$
		contains $q$ and no accepting states of $\A$
		(Lemma \ref{lemma:sub_dec=cover});
		\item
		Given a permutation \DFA $\A$ and a rejecting state $q$, 
		we can determine the existence of
		an orbit-\DFA covering $q$ in non-deterministic time
		$\mathcal{O}(|\A|^2)$, and in deterministic time
		$\mathcal{O}(2^kk \cdot|\A|)$,
		where $k$ is the number of rejecting states
		of $\A$ (Lemma~\ref{lemma:cover_dec_FPT}, (apx) Algorithm~\ref{alg:NP-permDFA-Decomp}).
	\end{itemize}
	These results directly imply Theorem~\ref{thm:permutation_automata_prime_NP}.
	We also apply them to
	show that the \Decomp
	problem is trivial for permutation \DFAs
	with a prime number of states.
	
	\begin{restatable}{theorem}{primeIsPrime}
		\label{theorem:prime_is_prime}
		Let $\A$ be a permutation \DFA
		with at least one accepting state
		and one rejecting state.
		If the number of states of $\A$ is prime, then $\A$ is prime.
	\end{restatable}

	\subsection{Proof of Theorem \ref{thm:permutation_automata_prime_NP}}

\begin{figure}[]
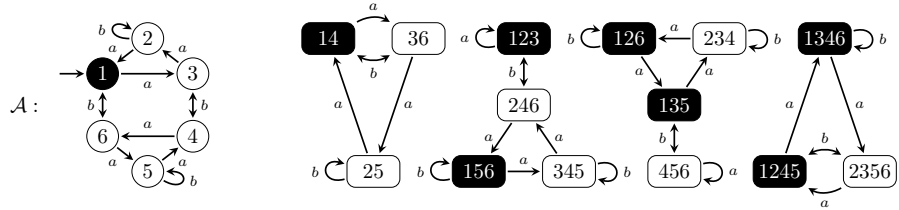

		\centering
		\scalebox{.8}{\drawOrbits}
		\caption{A \DFA $\A$
			together with some of its orbit-\DFAs.
			Accepting states are depicted in black,
			an orbit-\DFA can be obtained
			by setting a subset containing
			a $1$ as an initial state.
			For instance the orbit-\DFAs
			$\A^{\{1,2,3\}}$ and $\A^{\{1,5,6\}}$
			form a decomposition of $\A$.
		} \label{ex:orbit}
	\end{figure}

	Consider a \DFA $\A=\zug{\Sigma,Q,q_I,\delta,F}$.
	We extend $\delta$ to subsets $U \subseteq Q$
	in the expected way:
	\[
	\delta(U,w) = \{q \in Q \st q = \delta(p,w) \textup{ for some } p \in U \}
	\textup{ for every word $w \in \Sigma^*$}.
	\]
	The \emph{orbit} of $U$ is the collection
	$\mathcal{C}_U = \{ \delta(U,w) \subseteq Q \st w \in \Sigma^*\}$
	of subsets of $Q$ that can be reached from $U$
	by the action of $\delta$.
	If the subset $U \subseteq Q$ contains the initial state $q_I$ of $\A$,
	we define the \emph{orbit-\DFA}
	$\A^{U}=\zug{\Sigma,\mathcal{C}_U,U,\delta,\mathcal{C}'}$,
	where the state space $\mathcal{C}_U$ is the orbit of $U$,
	and the set $\mathcal{C}'$ of accepting states is composed of the sets $U' \in \mathcal{C}_U$
	that contain at least one of the accepting states of $\A \colon U' \cap F \neq \varnothing$.
	Note that $\A^{U}$ can alternatively be defined as the standard subset construction
	starting with the set $U \subseteq Q$ as initial state.
	The definition of the accepting states guarantees that $L(\A) \subseteq  L(\A^U)$:
	
	%@MAGICAPX
	\begin{restatable}[\appendixProof]{proposition}{orbitSmaller}\label{prop:orbit=smaller}
		%\begin{restatable}{proposition}{orbitSmaller}\label{prop:orbit=smaller}
		Every orbit-\DFA $\A^U$ of a \DFA $\A$ satisfies $L(\A) \subseteq L(\A^U)$.
	\end{restatable}
	\begin{toappendix}
		\orbitSmaller*
		\begin{proof}
			For every word $w$ accepted by $\A$,
			the state $\delta(q_I,w)$ that $\A$ visits after reading $w$
			is accepting.
			Moreover, as the initial state $q_I$ of $\A$ is in $U$,
			the state $\delta(U,w)$ that $\A^U$ visits after reading $w$
			contains the state $\delta(q_I,w) \in F$.
			Therefore, we get that $\delta(q_I,w) \in \delta(U,w) \cap F$,
			hence $\delta(U,w)$ is an accepting state of $\A^U$,
			which proves that $w \in L(\A^U)$.
		\end{proof}
	\end{toappendix}
	%\past{
	%	\begin{lemma}
	%		Every composite permutation \DFA can be decomposed into its orbit-\DFAs.
	%	\end{lemma}
	%}
	%
	%\begin{lemma}\label{lemma:permutation_orbit_decomposition}
	%	Let $\A=\zug{\Sigma,P,p_I,\delta_\A,F_\A}$ be a composite permutation \DFA and let $\B_1,\B_2, \ldots, \B_n$ be a decomposition of $\A$
	%	Then, for every $1 \leq i \leq n$, there exists an orbit-\DFA $\A^{U_i}$ such that $|\A^{U_i}| \leq |\B_i|$ and $L(\A) \subseteq L(\A^{U_i}) \subseteq L(\B_i)$ and  $\A^{U_1},\A^{U_2}, \ldots, \A^{U_n}$ is a decomposition of $\A$ into orbit-\DFAs.
	%\end{lemma}
	%
	\boldtitle{Example}
	Let us detail the orbits of the
	\DFA $\A$ depicted in Figure \ref{ex:orbit}.
	This \DFA contains six states,
	and generates the following non-trivial orbits
	on its subsets of states:
	\begin{itemize}
		\item
		The $15$ subsets of size $2$
		are split into two orbits:
		one of size $3$,
		and one of size $12$;
		\item 
		The $20$ subsets of size $3$
		are split into three orbits:
		two of size $4$,
		and one of size $12$;
		\item
		The $15$ subsets of size $4$
		are split into two orbits,
		one of size $3$,
		and one of size $12$.
	\end{itemize}
	Figure \ref{ex:orbit}
	illustrates the four orbits smaller than $|\A|$:
	they induce seven orbit-\DFAs,
	obtained by setting as initial state
	one of the depicted subsets
	containing the initial state $1$ of $\A$.
	
	\medskip
	\noindent
	In order to prove that a \DFA is composite
	if and only if it can be decomposed
	into its orbit-\DFAs,
	we prove that every factor $\B$
	of a permutation \DFA $\A$ can be turned into
	an orbit-\DFA $\A^U$ that is also a factor of $\A$,
	and satisfies $L(\A^U) \subseteq L(\B)$.
	Our proof is based on a 
	known result stating that factors
	can be turned into
	permutation \DFAs:
	
	\begin{lemma}[{\cite[Theorem 7.4]{KM15}}]\label{lemma:permutation_perm_decomposition}
		Let $\A$ be a permutation \DFA.
		For every factor $\B$ of $\A$,
		there exists a permutation \DFA $\C$
		satisfying $|\C|\leq|\B|$ and $L(\A) \subseteq L(\C)\subseteq L(\B)$.
	\end{lemma}
	We strengthen this result
	by showing how to transform
	factors into orbit-\DFAs:
	
	\begin{lemma}\label{lemma:permutation_orbit_decomposition}
		Let $\A$ be a permutation \DFA.
		For every factor $\B$ of $\A$,
		there exists an orbit-\DFA $\A^U$ of $A$
		satisfying $|\A^U|\leq|\B|$ and $L(\A) \subseteq L(\A^U)\subseteq L(\B)$.
	\end{lemma}

	\begin{proof}
		Let $\A=\zug{\Sigma,Q,q_I,\delta,F}$
		be a permutation \DFA,
		and let $\B$ be a factor of $\A$.
		By Lemma \ref{lemma:permutation_perm_decomposition},
		there exists a \emph{permutation}
		\DFA $\B'=\zug{\Sigma,S,s_I,\eta,G}$ satisfying $|\B'| \leq |\B|$ and
		$L(\A) \subseteq L(\B') \subseteq L(\B)$.
		We build,
		based on $\B'$, an orbit-\DFA $\A^U$ of $\A$ satisfying the statement.	
		
		We say that a state $q \in Q$ of $\A$ is \emph{linked} to a state $s \in S$ of $\B'$,
		denoted $q \sim s$,
		if there exists a word $u \in \Sigma^*$ satisfying $\delta(q_I,u) = q$
		and $\eta(s_I,u) = s$.
		Let $f : S \rightarrow 2^Q$
		be the function mapping every state $s \in S$
		to the set $f(s) \subseteq Q$ containing all
		the states $q \in Q$ that are linked to $s$ (i.e.\ satisfying $q \sim s$).
		We set $U = f(s_I)$.
		In particular, the initial state $q_I$ of $\A$ is in $U$
		since $\delta(q_I,\varepsilon) = q_I$ and $\eta(s_I,\varepsilon) = s_I$.
		We show that the orbit-\DFA $\A^{U}$ satisfies the desired conditions:
		$|\A^{U}| \leq |\B'|$
		and
		$L(\A) \subseteq L(\A^{U}) \subseteq L(\B')$.
		
		First, we show that $|\A^{U}| \leq |\B'|$
		by proving that the function $f$ defined earlier maps $S$ surjectively into
		the orbit of $U$, which is the state space of $\A^{U}$.
		Since both $\A$ and $\B'$ are permutation \DFAs, we get that for all $q \in Q$, $s \in S$ and $a \in \Sigma$, then $q \sim s$ if and only if $\delta(q,a) \sim \eta(s,a)$ holds.\footnote{Remark that for general \DFAs we only get that $q \sim s$ implies $\delta(q,a) \sim \eta(s,a)$ from the determinism. It is the \emph{backward determinism} of the permutation \DFAs $\A$ and $\B'$ that gives us the reverse implication.}
		Therefore, for every word $v \in \Sigma^*$,
		$f(\eta(s_I,v)) = \delta(f(s_I),v) = \delta(U,v)$.
		This shows that, as required, the image of the function $f$ is the orbit of $U$,
		and $f$ is surjective.
		
		To conclude, we show that $L(\A) \subseteq L(\A^{U}) \subseteq L(\B')$.
		Proposition \ref{prop:orbit=smaller} immediately
		implies that $L(\A) \subseteq L(\A^{U})$.
		Therefore it is enough to show that $L(\A^{U}) \subseteq L(\B')$.
		Let $v \in L(\A^{U})$.
		By definition of an orbit-\DFA, this means that
		the set $\delta(U,v)$ contains an accepting state $q_F$ of $\A$.
		Since, as stated earlier, $f(\eta(s_I,v)) = \delta(U,v)$,
		this implies (by definition of the function $f$)
		that the accepting state $q_F$ of $\A$ is linked to $\eta(s_I,v)$,
		i.e., there exists a word $v' \in \Sigma^*$ such that $\delta(q_I,v') = q_F$
		and $\eta(s_I,v') = \eta(s_I,v)$.
		Then $\delta(q_I,v') = q_F$ implies that $v'$ is in the language of $\A$.
		Moreover, since $L(\A) \subseteq L(\B')$ by supposition,
		$v'$ is also accepted by $\B'$, i.e., $\eta(s_I,v')$ is an accepting state of $\B'$.
		Therefore, since $\eta(q_I,v') = \eta(q_I,v)$,
		the word $v$ is also in the language of $\B'$.
		This shows that $L(\A^{U}) \subseteq L(\B')$, which concludes the proof.
	\end{proof}
	
	As an immediate corollary,
	every decomposition
	of a permutation \DFA can be transformed,
	factor after factor,
	into a decomposition into orbit-\DFAs.
	
	\begin{corollary}\label{cor:permutation_orbit_decomposition}
		A permutation \DFA is composite
		if and only if it can be decomposed
		into its orbit-\DFAs.
	\end{corollary}
	
	\boldtitle{Orbit cover}
	Given a rejecting state $q \in Q \setminus F$ of $\A$,
	we say that the orbit-\DFA $\A^U$
	\emph{covers} $q$
	if $|\A^U| < |\A|$,
	and $\A^U$ contains a rejecting state $U' \subseteq Q$
	that contains $q$. 
	Remember that, by definition, this means that $U'$ contains no accepting state of $\A$,
	i.e., $U' \cap F = \varnothing$.
	We show that permutation \DFAs that can be decomposed into their orbit-\DFAs
	are characterized by the existence of orbit-\DFAs covering each of their rejecting states.
	
	\begin{lemma}\label{lemma:sub_dec=cover}
		A permutation \DFA $\A$ is decomposable into its orbit-\DFAs if and only if every
		rejecting state of $\A$
		is covered by an orbit-\DFA
		$\A'$ of $\A$ satisfying $|\A'| < |\A|$.
	\end{lemma}
	
	\begin{proof}
		Let $\A=\zug{\Sigma,Q,q_I,\delta,F}$ be a  permutation \DFA.
		We prove both implications.
		
		Suppose that $\A$ can be decomposed into its orbit-\DFAs $(\A^{U_i})_{1 \leq i \leq k}$,
		and let $q \in Q \setminus F$ be a rejecting state of $\A$.
		We show that $q$ is covered by every orbit-\DFA $\A^{U_i}$ that rejects a word $w \in \Sigma^*$ satisfying $\delta(q_I,w) = q$.
		Formally, let $w \in \Sigma^*$ be a word satisfying $\delta(q_I,w) = q$.
		Then $w \notin L(\A) = \bigcap_{i=1}^{n}L(\A^{U_i})$,
		hence there exists $1 \leq i \leq n$ such that $w \notin L(\A^{U_i})$. 
		Let $U' \subseteq Q$ be the state visited by $\A^{U_i}$ after reading $w$.
		Then, by applying the definition of an orbit-\DFA,
		we get that $q \in U'$ since $\delta(q_I,w) = q$,
		and $U' \cap F = \varnothing$ since $U'$ is a rejecting state of $\A^{U_i}$ (as $w \notin L(\A^{U_i})$).
		Therefore, $\A^{U_i}$ covers $q$.
		Moreover, $|\A^{U_i}| < |\A|$
		since $\A^{U_i}$
		is a factor of $\A$.
		
		Conversely, let us fix an enumeration
		$q_1,q_2, \ldots, q_m$
		of the rejecting states of $\A$,
		and suppose that for all $1 \leq i \leq m$
		there is an orbit-\DFA $\A^{U_i}$ of $\A$ 
		that covers $q_i$ and satisfies $|\A^{U_i}|<|\A|$.
		Let $(U_{i.j})_{1 \leq j \leq n_i}$ be an enumeration of
		the subsets in the orbit of $U_i$ that contain the initial state $q_I$ of $\A$.
		We conclude the proof by showing that $S = \{\A^{U_{i.j}} \st 1 \leq i \leq m, 1 \leq j \leq n_i\}$
		is a decomposition of $\A$.
		Note that we immediately get $|\A^{U_{i.j}}| = |\A^{U_i}| <|\A|$ for all $1 \leq i \leq m$ and $1 \leq j \leq n_i$.
		Moreover, Proposition \ref{prop:orbit=smaller}  implies  $L(\A) \subseteq  \bigcap_{\A' \in S} L(\A')$.
		To complete the proof, we show that $\bigcap_{\A' \in S} L(\A') \subseteq L(\A)$.
		Let $w \in \Sigma^*$ be a word rejected by $\A$.
		To prove the desired inclusion,
		we show that there is a \DFA $\A' \in S$
		that rejects $w$.
		Since $w \notin L(\A)$, the run of $\A$ on $w$
		starting from the initial state
		ends in a rejecting state $q_i$, for some $1 \leq i \leq m$.
		By supposition the orbit-\DFA $\A^{U_i}$ covers $q_i$,
		hence the orbit of $U_i$ contains a set $U' \subseteq Q$ that contains $q_i$ and no accepting state.
		Note that there is no guarantee that $\A^{U_i}$ rejects $w$:
		while the set $\delta(U_i, w)$ contains $q_i$,
		it is not necessarily equal to $U'$,
		and might contain accepting states.
		However, as $\A$ is a permutation \DFA,
		we can reverse all of the transitions of $\A$
		to get a path labeled by the reverse of $w$ that starts from $U'$ (that contains $q_i$),
		and ends in one of the sets $U_{i.j}$ (that contains $q_I$).\footnote{
			Remark that, if $\A$ is not a permutation \DFA, then some states might not have incoming transitions for every letter.
			Thus, the reversal of $w$ might not be defined.
		}
		Therefore, by reversing this path back to normal,
		we get that $\delta(U_{i.j}, w) = U'$,
		hence the orbit-\DFA $\A^{U_{i.j}} \in S$ rejects $w$.
		Therefore, every word rejected by $\A$ is rejected by an orbit-\DFA $\A' \in S$,
		which shows that $\bigcap_{\A' \in S} L(\A') \subseteq L(\A)$.
	\end{proof}
	
	This powerful lemma allows us to easily
	determine whether a permutation \DFA is composite
	if we know its orbits.
	For instance, the \DFA $\A$ depicted in
	Figure \ref{ex:orbit}
	is composite since the orbit-\DFA $\A^{\{1,2,3\}}$ 
	covers its five rejecting states.
	Following the proof of
	Lemma~\ref{lemma:sub_dec=cover},
	we get that
	($\A^{\{1,2,3\}},\A^{\{1,5,6\}})$ is a decomposition of $\A$,
	and so is $(\A^{\{1,2,6\}},\A^{\{1,3,5\}})$. 
	
	To conclude, we give an algorithm
	checking if a rejecting
	state is covered by an orbit-\DFA.
	
	\begin{lemma}\label{lemma:cover_dec_FPT}
		Given a permutation \DFA $\A$ and a rejecting state $q$,
		we can determine the existence of an orbit-\DFA
		that covers $q$ in nondeterministic time
		$\mathcal{O}(k \cdot |\A|^2)$, and in deterministic time $\mathcal{O}(2^kk \cdot|\A|^2)$,
		where $k$ is the number of rejecting states of $\A$.
	\end{lemma}
	
	\begin{proof}
		We can decide in \NP whether there exists an orbit-\DFA $\A^{U}$ of $\A$
		that covers $p$:
		we non-deterministically guess among the set of rejecting states of $\A$ a subset $U'$ containing $p$. 
		Then, we check in polynomial time that the orbit of $U'$ is smaller than $|\A|$.
		This property can be checked in time $\mathcal{O}(|\A|^2)$.
		Since $\A$ is trim, in the orbit of $U'$ there is a set $U$ containing the initial state of $\A$.
		Moreover, since $\A$ is a permutation \DFA,  $U$ and $U'$ induce the same orbit.
		Hence, $p$ is covered by the orbit-\DFA $\A^{U}$.
		Finally, we can make this algorithm deterministic by searching through the $2^k$ possible subsets $U'$ of the set of rejecting states of $\A$.% instead of guessing one.
	\end{proof}
	
	\begin{toappendix}
		% ++++++++++++ IMPORT ALGO ++++++++++++
		%\begin{figure}[h]
		%\centering
		%\includegraphics{algo_dec_permut.png}
		%\stepcounter{theorem}
		%\end{figure}
		% ++++++++++++ ++++++++++++ +++++++++++
		\begin{algorithm}[]
			\showalgo{
				\DontPrintSemicolon
				\SetKwFunction{main}{\normalfont isComposite}
				\SetKwFunction{cover}{\normalfont cover}
				\SetKwComment{ccc}{\color{gray}/* }{\color{gray}\ */}
				\SetKwProg{fun}{Function}{}{end}
				\SetKw{true}{\normalfont True}
				\SetKw{false}{\normalfont False}
				\SetKw{with}{with}
				\SetKw{and}{and}	
				\SetKw{neg}{not}
				\SetKw{guess}{guess}
				\SetKw{coverForall}{\normalfont cover\_all}
				\SetKw{coverExists}{\normalfont cover\_current}
				
				\fun{\main{$\A = \zug{\Sigma, Q, q_I, \delta, F}\colon$\normalfont permutation \DFA}}{
					\ForEach{$p \in Q\setminus F$}{
						\guess $U$ \with $\{p\} \subseteq U \subseteq Q\setminus F$\ccc*[f]{\color{gray}guess rejecting state $U$ of some orbit-\DFA, such that $U$ contains rejecting state $p$ of $\A$}\;
						
						\lIf{$\neg\ \cover(\A, p, U)$}{\Return{\false}}
					}
					\Return{\true}
				}
				
				\BlankLine
				\fun{\cover{$\A = \zug{\Sigma, Q, q_I, \delta, F}\colon$\normalfont permutation \DFA, $p \in Q\setminus F$, $U \subseteq Q\setminus F$}}{
					$\mathcal{C}_U^\text{old} = \varnothing$\;
					$\mathcal{C}_U \affects \{U\}$\;
					\While{$\mathcal{C}_U \neq \mathcal{C}_U^\text{old}$ \and $|\mathcal{C}_U| < |Q|$}{
						$\mathcal{C}_U^\text{old} \affects \mathcal{C}_U$
						
						$\mathcal{C}_U \affects	\mathcal{C}_U \cup \{\delta(S, \sigma) \st S \in \mathcal{C}_U, \sigma \in \Sigma\}$
					}
					\lIf(\ccc*[f]{\color{gray}check that orbit-\DFA is factor}){$|\mathcal{C}_U| \geq |Q|$}{\Return{\false}}
					\ForEach{$S \in \mathcal{C}_U$}{
						\lIf(\ccc*[f]{\color{gray}check that $U$ is reachable from the inital state of the orbit-\DFA}){$q_I \in S$}{\Return{\true}}
					}
					\Return{\false}
				}
			}
			\caption{\NP-algorithm for the \Decomp problem for permutation \DFAs.}
			\label{alg:NP-permDFA-Decomp}
		\end{algorithm}
	\end{toappendix}

	\subsection{Proof of Theorem~\ref{theorem:prime_is_prime}}
	
	Thanks to the notion of orbit \DFAs we are able to prove that a permutation \DFA which has a prime number of states with at least one accepting and one rejecting, is prime.
	\begin{proof}
		Let $\A=\zug{\Sigma,Q,q_I,\delta,F}$
		be a trim permutation \DFA
		with a state space $Q$
		of prime size
		that contains at least one accepting state and one rejecting state.
		We show that the only orbit of $\A$
		smaller than $|Q|$
		is the trivial orbit $\{Q\}$.
		This implies that $\A$ cannot be decomposed into its orbit-\DFAs,
		which proves that $\A$ is prime by Lemma \ref{lemma:permutation_orbit_decomposition}.
		
		Let us consider a strict subset $U_1 \neq \varnothing$
		of the state space $Q$,
		together with its orbit $\mathcal{C}_{U_1} = \{U_1,U_2,\ldots,U_m\}$.
		We prove that $m \geq |Q|$.
		First, we show that all the $U_i$ have the same size: 
		since $U_i$ is an element of the orbit of $U_1$,
		there exists a word $u_i \in \Sigma^*$ satisfying $\delta(U_1,u_i)=U_i$,
		and, as every word in $\Sigma^*$
		induces via $\delta$ a permutation on the state space,
		$|U_i| = |\delta(U_1,u_i)| = |U_1|$.
		Second, for every $q \in Q$,
		we define the \emph{multiplicity} of $q$
		in $\mathcal{C}_{U_1}$
		as the number $\lambda(q) \in \mathbb{N}$
		of distinct elements of $\mathcal{C}_{U_1}$
		containing the state $q$.
		We show that all the states $q$ have the same multiplicity:
		since $\A$ is trim,
		there exists a word $u_q \in \Sigma^*$
		satisfying $\delta(q_I,u_q)=q$,
		hence $u_q$ induces via $\delta$ a bijection
		between the elements of $\mathcal{C}_{U_1}$ containing
		$q_I$ and those containing $q$,
		and
		$\lambda(q) = \lambda(\delta(q_I,u_q)) = \lambda(q_I)$.
		By combining these results, we obtain $m \cdot |U_1| = \Sigma_{i=1}^m |U_i| = \Sigma_{q \in Q} \lambda(q) = \lambda(q_I) \cdot |Q|$. 
		Therefore, as $|Q|$ is prime by supposition,
		either $m$ or $|U_1|$ is divisible by $|Q|$.
		However, $U_1 \subsetneq Q$, hence $|U_1| < |Q|$,
		which shows that $m$ is divisible by $|Q|$.
		In particular, we get $m \geq |Q|$,
		which concludes the~proof.
	\end{proof}
	\section{Decompositions of Commutative Permutation \DFAs}\label{sec:commutative}
	We now study \emph{commutative} permutation
	\DFAs :
	a \DFA $\A=\zug{\Sigma,Q,q_I,\delta,F}$
	is commutative if 
	$\delta(q,uv) = \delta(q,vu)$
	for every state $q$ 
	and every pair of words $u,v \in \Sigma^*$.
	%Note that this implies that
	%for each word $w=w_1w_2\dots w_n\in \Sigma^*$, $w \in L(\A)$ iff $w_{\pi(1)}w_{\pi(2)}\dots w_{\pi(n)} \in L(\A)$ for every permutation $\pi \colon \{1, 2,\dots, n\} \to \{1, 2, \dots, n\}$. Hence, for commutative \DFAs the acceptance of a word only depends on the number of occurrences of each letter.
	%Commutative permutation \DFAs
	%form a strict subclass of the \emph{normal}
	%permutation \DFAs studied in~\cite{KM15}.
	Our main contribution is an \NLOGSPACE algorithm for the \Decomp problem
	for commutative permutation \DFAs.
	Moreover, we show that the complexity
	goes down to \LOGSPACE
	for alphabets of fixed size.
	
	\begin{theorem}\label{thm:commutative_automata_prime_LS}
		The \Decomp problem
		for commutative permutation \DFAs
		is in \NLOGSPACE,
		and in \LOGSPACE when the size of the alphabet is fixed.
	\end{theorem}
	\noindent
	The proof of Theorem \ref{thm:commutative_automata_prime_LS}
	is based on the notion of \emph{covering word}:
	a word $w \in \Sigma^*$
	covers a rejecting state $q$ of
	a \DFA $\A = \zug{\Sigma, Q,q_I,\delta,F}$
	if $\delta(q,w) \neq q$,
	and for every $\lambda \in \N$,
	the state $\delta(q,w^\lambda)$
	is rejecting.
	We prove two related key results:
	\begin{itemize}
		\item 
		A commutative permutation \DFA
		is composite if and only if
		each of its rejecting states
		is covered by a word (Lemma \ref{lemma:Newword_dec=cover}).
		\item
		We can decide in \NLOGSPACE
		(\LOGSPACE when the size of the alphabet is fixed)
		if a given rejecting state of a \DFA
		is covered by a word
		(Lemma \ref{lemma:com-perm-decomp-LOGSPACE}, and Algorithm~\ref{alg:decomp-Log} in appendix)
		%	\commentApx{(Lemma \ref{lemma:com-perm-decomp-LOGSPACE}, and Algorithm~\ref{alg:decomp-Log} in appendix)}{(\textcolor{red}{TOTO})}.
		%@MAGICAPX
	\end{itemize}
	These results immediately imply
	Theorem \ref{thm:commutative_automata_prime_LS}.
	%\emph{quotients}
	%of a \DFA, a specific kind of orbit-\DFAs.
	%We show that a permutation \DFA
	%is composite if and only if
	%each of its rejecting state is \emph{covered}
	%by a quotient \DFA
	%(Lemma~\ref{lemma:word_dec=cover}),
	%which is decidable in \LOGSPACE
	%(Lemma~\ref{lemma:commutative_automata_quotient_dec_LS}).
	We conclude this section by showing an upper bound
	on the width and constructing
	a family of \DFAs of polynomial width.
	
	\begin{theorem}\label{theorem:poly_family}
		The width of every composite permutation \DFA is smaller than its size.
		Moreover, for all $m,n \in \N$
		such that $n$ is prime,
		there exists a commutative permutation \DFA
		of size $n^m$ and width $(n-1)^{m-1}$.
	\end{theorem}
	%To prove Theorem \ref{theorem:poly_family},
	We show that the width of a commutative permutation \DFA
	is bounded by its number of rejecting states (Lemma \ref{lemma:Newword_dec=cover}).
	Then, for each $m,n \in \mathbb{N}$
	with $n$ prime,
	we define a \DFA $\A_n^m$
	of size $n^m$
	that can be decomposed into $(n-1)^{m-1}$ factors
	(Proposition \ref{prop:composite}),
	but not into $(n-1)^{m-1}-1$
	(Proposition \ref{prop:not_composite}).

	\subsection{Proof of Theorem \ref{thm:commutative_automata_prime_LS}}
	
	The
	%\petra{TODO $u$ vs $w$ fr om here on}
	proof is based on the following key property
	of commutative permutation \DFAs: 
	In a permutation \DFA $\A$,
	every input word acts as a permutation
	on the set of states, 
	generating disjoint cycles,
	and if $\A$ is commutative
	these cycles form an orbit.
	
	\begin{restatable}{proposition}{propDisjoint}\label{prop:disjoint}
		Let $\A=\zug{\Sigma,Q,q_I,\delta,F}$ be a commutative permutation \DFA.
		For all $u \in \Sigma^*$, the sets $(\{\delta(q,u^{\lambda}) \st \lambda \in \N\})_{q \in Q}$ partition $Q$ and form an orbit of $\A$.
	\end{restatable}
	\begin{proof}
		Let $\A=\zug{\Sigma,Q,q_I,\delta,F}$ be a commutative permutation \DFA.
		Given $u \in  \Sigma^*$ and $q \in Q$, the sequence of states $\delta(q,u), \delta(q,u^2), \dots, \delta(q,u^i)$ visited by applying  $\delta$ on iterations of $u$ eventually repeats i.e.\ $\delta(q,u^x) = \delta(q,u^y) = p$ for some $x, y \in \N$ and $p\in Q$.
		Since $\A$ is a permutation \DFA, it is both forward and backward deterministic, thus the set of visited states $\{\delta(q,u^{\lambda}) \st \lambda \in \N\}$ is a cycle that contain both $p$ and $q$.
		The collection $(\{\delta(q,u^{\lambda}) \st \lambda \in \N\})_{q \in Q}$ forms an orbit of $\A$ by commutativity.
		Formally, for all  $u, v \in \Sigma^*$ and every $q \in Q$, we have:
		$
		\delta(\{\delta(q,u^{\lambda}) | \lambda \in \N\}, v)
		= \{\delta(q,u^{\lambda}v) | \lambda \in \N\}
		= \{\delta(q,vu^{\lambda}) | \lambda \in \N\}
		= \{\delta(\delta(q,v),u^{\lambda}) | \lambda \in \N\}$.\hspace{-2pt}
	\end{proof}

	We proved with Corollary \ref{cor:permutation_orbit_decomposition}
	and Lemma \ref{lemma:sub_dec=cover}
	that a permutation \DFA is composite
	if and only if each of its rejecting
	states is covered by an orbit-\DFA.
	We now reinforce this result
	for \emph{commutative} permutation \DFAs.
	As stated before, we say that a word $u \in \Sigma^*$
	\emph{covers} a rejecting state $q$ of
	a \DFA $\A = \zug{\Sigma, Q,q_I,\delta,F}$
	if $u$ induces from $q$ a non-trivial cycle
	composed of rejecting states:
	$\delta(q,u) \neq q$,
	and $\delta(q,u^\lambda)$
	is rejecting for all $\lambda \in \N$.
	Note that the collection $(\{\delta(q,u^\lambda) \st \lambda \in \N\})_{q \in Q}$ forms an orbit of $\A$ by Proposition~\ref{prop:disjoint}.
	We show that we can determine if $\A$ is composite
	by looking for words covering its rejecting states.

	\begin{restatable}{lemma}{wordCover}\label{lemma:Newword_dec=cover}
		For every $k \in \mathbb{N}$,
		a commutative permutation \DFA $\A$ 
		is $k$-factor composite
		if and only if there exist $k$ words that,
		together, cover all the rejecting states of $\A$.
	\end{restatable}
	
	% ----- SKETCH -----
	%\begin{proof}[Proof sketch]
	%	Let $\A$ be a \DFA.
	%	The proof follows the same ideas as the proof of Lemma \ref{lemma:sub_dec=cover}.
	%	Transforming a set of covering words into a decomposition is easy:
	%	By proposition \ref{prop:disjoint}
	%	every word induces an orbit-\DFA,
	%	and we show that if we have $k$ words that, together,
	%	cover all the rejecting states of $\A$,
	%	then the $k$ induced orbit-\DFAs form a decomposition of $\A$.
	%	To prove the converse direction, we suppose that $\A$ has a $k$-factor decomposition $(\B_i)_{1 \leq i \leq k}$,
	%	and we use it to create a set of $k$ words that cover all the rejecting states.
	%	To this end, we first apply Lemma \ref{lemma:permutation_perm_decomposition}
	%	to transform the decomposition of $\A$ into a decomposition $(\C_i)_{1 \leq i \leq k}$
	%	composed of \emph{permutation} factors.
	%	For every $1 \leq i \leq k$, we extract a word $u_i$
	%	from $\C_i$ as follows:
	%	using the fact that $|\C_i| < |\A|$,
	%	we build a word $u_i$
	%	that moves the initial state of $\A$
	%	but fixes the initial state of $\C_i$.
	%	Using this key property and the fact that $(\C_i)_{1 \leq i \leq k}$ is a decomposition of $\A$ allows us to show
	%	that the words $u_1,u_2,\ldots,u_k$ cover all the rejecting states.
	%\end{proof}
	%\begin{toappendix}
	%\wordCover*
	\begin{proof}
		Let $\A=\zug{\Sigma,Q,q_I,\delta,F}$ be a commutative permutation \DFA and $k \in \N$.
		We start by constructing $k$ factors based on $k$ covering words.
		Suppose that there exist
		$k$ words $u_1,u_2,\ldots,u_k$
		such that every rejecting state $q \in Q \setminus F$
		is covered by one of the $u_i$.
		Note that all the $u_i$ covering at least
		one state $q$ do not act as the identity
		on $Q$ (since $\delta(q,u_i) \neq q$),
		therefore we suppose, without loss of generality,
		that none of the $u_i$ acts as the identity
		on $Q$.
		For every $1 \leq i \leq k$,
		let $U_i = \{ \delta(q_I,u_i^\lambda) \st \lambda \in \N \}$.
		We show that
		$(\A^{U_i})_{1 \leq i \leq k}$
		is a decomposition of $\A$.
		As none of the $u_i$ acts as the identity on $Q$, Proposition \ref{prop:disjoint}
		implies that every $\A^{U_i}$
		is smaller than $\A$.
		Moreover,
		Proposition \ref{prop:orbit=smaller}
		implies that
		$L(\A) \subseteq L(\A^{U_i})$,
		hence
		$L(\A) \subseteq \bigcap_{j=1}^kL(\A^{U_j})$.
		To conclude, we show that
		$\bigcap_{j=1}^kL(\A^{U_j}) \subseteq L(\A)$.
		Let $u \notin L(\A)$.
		By supposition, there exists $1 \leq i \leq k$
		such that $u_i$ covers $\delta(q_I,u)$.
		As a consequence, the set
		\[\begin{array}{lll}
			\delta(U_i,u) &=&
			\delta(\{ \delta(q_I,u_i^\lambda) \st \lambda \in \N \},u) =
			\{ \delta(q_I,u_i^\lambda u) \st \lambda \in \N \} =
			\{ \delta(q_I,u u_i^\lambda) \st \lambda \in \N \}\\
			&=&
			\{ \delta(\delta(q_I,u),u_i^\lambda) \st \lambda \in \N \}
		\end{array}
		\]
		contains no accepting state of $\A$,
		hence it is a rejecting state of $\A^{U_i}$.
		As a consequence, 
		we get $u \notin L(\A^{U_i}) \supseteq \bigcap_{j=1}^kL(\A^{U_j})$, which proves that
		$\bigcap_{j=1}^kL(\A^{U_j}) \subseteq L(\A)$.
		
		We now construct $k$ covering words based on $k$ factors.
		Suppose that $\A$ has a $k$-factor decomposition
		$(\B_i)_{1 \leq i \leq k}$.
		Lemma \ref{lemma:permutation_perm_decomposition}
		directly implies that this decomposition
		can be transformed into a decomposition
		$(\C_i)_{1 \leq i \leq k}$
		of $\A$, where
		$\C_i = \zug{\Sigma,S_i,s^i_I,\eta_i,G_i}$
		are permutation \DFAs.
		For every $1 \leq i \leq k$,
		we build a word $u_i$ based on $\C_i$,
		we prove that every
		rejecting state of $\A$
		is covered by one of these $u_i$.
		Consider $1 \leq i \leq k$.
		Since $\C_i$ is a factor of $\A$,
		in particular $|\C_i|<|\A|$,
		hence
		there exist two input words $v_i,w_i \in \Sigma^*$
		such that $\A$ reaches different states on $v_i$ and $w_i$,
		but $\C_i$ reaches the same state:
		$\delta(q_I,v_i) \neq \delta(q_I,w_i)$
		but $\eta_i(s^i_I,v_i) = \eta_i(s^i_I,w_i)$.
		Note that both $\A$ and $\C_i$ are permutation
		\DFAs,
		hence there exists a power $v_i^{\kappa_i}$
		of $v_i$
		that induces the identity function
		on both state spaces $Q$ and $S_i$.
		We set $u_i  = w_iv_i^{\kappa_i-1}$,
		which guarantees that:
		\[
		\begin{array}{l}
			\delta(q_I,u_i)
			= \delta(\delta(q_I,w_i),v_i^{\kappa_i-1})
			\neq \delta(\delta(q_I,v_i),v_i^{\kappa_i-1})
			= \delta(q_I,v_i^{\kappa_i}) = q_I;
			\\
			\eta_i(s_I^i,u_i) = \eta_i(\eta_i(s_I^i,w_i),v_i^{\kappa_i-1})
			= \eta_i(\eta_i(s_I^i,v_i),v_i^{\kappa_i-1}) = \eta_i(s_I^i,v_i^{\kappa_i}) = s_I^i.
		\end{array}
		\]
		In other words, $u_i$ moves the initial state $q_I$ of $\A$,
		but fixes the initial state $s_I^i$ of $\C_i$.
		
		We now prove that each rejecting state of $\A$ is covered by one of the $u_i$.
		Let $q \in Q \setminus F$ be a
		rejecting state of $\A$.
		Since $\A$ is trim,
		there exists a word $u_q \in \Sigma^*$
		such that $\delta(q_I,u_q) = q$.
		Then, as $u_q \notin L(\A)$
		and $(\C_i)_{1 \leq i \leq k}$
		is a decomposition of $\A$,
		there exists $1 \leq i \leq k$
		such that $u_q \notin L(\C_i)$.
		We show that the word $u_i$ 
		covers the rejecting state $q$:
		we prove that $\delta(q,u_i) \neq q$,
		and that $\delta(q,u_i^\lambda)$
		is rejecting for every $\lambda \in \N$.
		First, since $\A$ is a commutative permutation $\DFA$
		and $u_i$ moves $q_I$,
		we get that $\delta(q,u_i) = \delta(q_I,u_qu_i) = \delta(q_I,u_iu_q) \neq \delta(q_I,u_q) = q$.
		Moreover,
		for all $\lambda \in \mathbb{N}$,
		Since $u_q \notin L(\C_i)$ by supposition and $u_i$ fixes $s_I^i$,
		the \DFA $\C_i$ also rejects the word $u_i^{\lambda}u_q$.
		Therefore, as $L(\A) \subseteq L(\C_i)$,
		we finally get that
		$\delta(q,u_i^{\lambda}) = \delta(q_I,u_qu_i^{\lambda}) = \delta(q_I,u_i^{\lambda}u_q)$
		is a rejecting state of $\A$.
	\end{proof}

	By Lemma \ref{lemma:Newword_dec=cover},
	to conclude the proof
	of Theorem \ref{thm:commutative_automata_prime_LS}
	we show that
	we can decide in \NLOGSPACE
	(and in \LOGSPACE when
	the size of the alphabet is fixed)
	whether  a given rejecting state of a \DFA
	is covered by a word
	(since in the \Decomp problem we can afford to pick a covering word for each state).
	As we consider commutative permutation \DFAs, we can represent a covering word by the number of occurrences of each letter,
	which are all bounded by $|Q|$. 
	
	%@MAGICAPX
	\begin{restatable}[\appendixProof]{lemma}{lemmaLogSpace}\label{lemma:com-perm-decomp-LOGSPACE}
		%\begin{restatable}{lemma}{lemmaLogSpace}\label{lemma:com-perm-decomp-LOGSPACE}
		Let $\A$ be a commutative permutation \DFA and $p$ a rejecting state.
		\begin{enumerate}
			\item\label{item:Logspace} We can determine the existence of a word covering $p$ in space $\mathcal{O}(|\Sigma|\cdot \log{|Q|})$;
			\item\label{item:NL} We can determine the existence of a word covering $p$ in \NLOGSPACE;
		\end{enumerate}
	\end{restatable}
	
	\begin{toappendix}
		\lemmaLogSpace*
		% \begin{lemma}
		% 	\label{lemma:com-perm-decomp-LOGSPACE2}
		% %	%	Primality of permutation commutative is in \LOGSPACE.
		% %	For a fixed alphabet size, the \textsc{Decomposition} problem for commutative permutation \DFAs is in \LOGSPACE.\todo[color=magenta]{Should we reformulate the statement by omitting the loop over $p$}
		% %	
		% 	Given a commutative permutation \DFA $\A$ and a rejecting state $p$, we can determine the existence of a word covering $p$ in space $\mathcal{O}(|\Sigma|\cdot \log{|Q|})$.
		% \end{lemma}
		\begin{proof}[Proof of Item \ref{item:Logspace}]
			%	Let $\A=\zug{\Sigma,Q,q_I,\delta,F}$ be a commutative permutation \DFA. As the class of normal \DFAs contains the class of commutative permutation \DFAs we get with \red{Lemma~\ref{lem:quotient:orna}} that $\A$ is composite if and only if it is decomposable into its partition-quotient \DFAs. \todo[color=magenta]{Adapt this s.t. it uses the new statements above}
			%	In order to check the latter, we need to find for every rejecting state $p$ of $\A$, a partition-quotient \DFA that covers $p$. As the set of states of a partition-quotient $\A^U=\zug{\Sigma,\mathcal{C}_U,U,\delta,\mathcal{C'}}$ of $\A$ forms a partition on the state set $Q$, (*) a state $S \in \mathcal{C}_U$ cannot contain two states $p, q \in Q$ simultaneously for which for some letter $\sigma$,  $\delta(p,\sigma) \in S$ and $\delta(q, \sigma) \notin S$.
			%	This observation (*) will help us to compute for a rejecting state $p$, some set $S$ such that the partition quotient $\A^U$ contains $S$ as a state and covers $p$.
			
			Let $\A=\zug{\Sigma,Q,q_I,\delta,F}$ be a commutative permutation \DFA with alphabet $\Sigma = \{\sigma_1, \sigma_2, \dots, \sigma_m\}$.
			Note that 
			%	for a trim commutative permutation \DFA  over the alphabet $\Sigma = \{\sigma_1, \sigma_2, \dots, \sigma_m\}$, it holds that 
			(*) 
			for every pair of states $p, q\in Q$, there exists a word $w_{p,q} = \sigma_1^{i_1}\sigma_2^{i_2}\dots \sigma_m^{i_m} \in \Sigma^*$ with $i_1 + i_2 + \dots + i_m < |Q|$ such that $\delta(p, w_{p,q}) = q$. Further, note that for commutative permutation \DFAs, $\delta(p, w^{|Q|}) = p$ for every state $p$ and word $w\in \Sigma^*$.
			Let $p \in Q\setminus F$ be a rejecting state of $\A$.
			We can decide in logarithmic space whether there exists a word covering $p$ as follows. 
			Pick a state $q\in Q$ with $p \neq q$ and determine $w_{p,q}$
			%	For two states $p, q$ we can find the word $w_{p, q}$ 
			by iterating over all $|Q|^{|\Sigma|}$ words of the form described in (*) and checking whether $\delta(p, w_{p, q}) = q$. Due to observation (*), we can represent each potential candidate for $w_{p,q}$ with $|\Sigma|\cdot\log(|Q|)$ bits.
			
			Next, we have to ensure that all states in the cycle induced by $w_{p,q}$ on $p$ are rejecting, i.e., we have to check that $\delta(p, w_{p,q}^\lambda) \notin F$ for all $\lambda\leq |Q|$. 
			%	
			%	
			%%%%%%%
			%	The index of the currently considered word can hence be stored in $|\Sigma|\cdot\log(|Q|)$ bits. 
			%%%%%%%
			%	
			%	
			%	We see $q$ as being in the same state $S \in \mathcal{C}_U$ as $p$. Since $\delta(p, w_{p,q}) = q$, observation (-) gives us, that the state $\delta(q, w_{p, q})$ must also be seen as being contained in $S$. We don't have to store the set $S$ (of size up to $|Q|$), we only have to ensure that for each state in $S$ its image under $w_{p, q}$ is contained in $S$ and that $S$ contains no accepting state. 
			Therefore, we use two pointers, one for the state $p$ and one for the image of $p$ under $w_{p,q}^\lambda$ where $\lambda$ denotes the current iteration step. As long as $\delta(p, w_{p,q}^\lambda)$ is rejecting and $\neq p$ we continue to compute  $\delta(p, w_{p,q}^{\lambda+1})$. If $\delta(p, w_{p,q}^\lambda)$ is accepting, we abort the computation and repeat the computation with some other state $q'$ instead of $q$. If we find $\delta(p, w_{p,q}^\lambda) = p$, we confirmed the existence of a word covering $p$. 
			%
			%	
			%	The word $w_{p,j}$ can be stored using $n$ numbers $\leq |Q|$ which can be encoded into one number of size?? 
			%	
			The current iteration step over the states $p$ and $q$ can be stored via two pointers. Note that the iteration step $\lambda$ don't have to be stored due to the permutation property which ensures us to encounter $p$ again, finally. 
			%	\todo[inline]{Where do we need the commutativity?}
		\end{proof}

		\begin{algorithm}
			\showalgo{
				\DontPrintSemicolon
				\SetKwFunction{main}{\normalfont isComposite}
				\SetKwFunction{mimic}{\normalfont mimic}
				\SetKwFunction{cover}{\normalfont cover}
				\SetKwComment{ccc}{\color{gray}/* }{\color{gray}\ */}
				\SetKwProg{fun}{Function}{}{end}
				\SetKw{true}{\normalfont True}
				\SetKw{false}{\normalfont False}
				\SetKw{with}{with}
				\SetKw{and}{and}
				\SetKw{guess}{guess}
				\SetKw{coverForall}{\normalfont cover\_all}
				\SetKw{coverExists}{\normalfont cover\_found}
				\SetKw{neg}{not}

				\fun{\main{$\A = \zug{\Sigma, Q, q_I, \delta, F}\colon$\normalfont commutative permutation \DFA}}{
					\ForEach{$p \in Q\setminus F$}{
						\coverExists \affects \false\;
						\ForEach{$q\in Q\setminus F$ \with $q \neq p$}{
							\lIf
							(\ccc*[f]{\color{gray}covering $p$ with $w_{p,q}$})
							{\cover{$\A, p, q$}}{\coverExists\affects \true}
						}
						\lIf(\ccc*[f]{\color{gray}no cover found for $p$}){\neg \coverExists}{\Return{\false}}
					}
					\Return{\true}\ccc*[f]{\color{gray}all state $p$ are covered}
				}
				\BlankLine
				\fun{\cover{$\A = \zug{\Sigma, Q, q_I, \delta, F}\colon$\normalfont commutative permutation \DFA, $p, q \in Q\setminus F$}}{
					$s \affects q$\;
					\While
					(\ccc*[f]{\color{gray}eventually $s = p$ $\A$ is a permuation \DFA})
					{$s \neq p$}{
						$s \affects$ \mimic{$p, q, s$}
						\ccc*[r]{\color{gray}thus $s \affects \delta(s, w_{p,q})$}
						\lIf
						(\ccc*[f]{\color{gray}contradiction of covering})
						{$s \in F$}{\Return{\false}}
					}
					\Return{\true}\ccc*[f]{\color{gray}encountered $p$ again without hitting state in $F$}
				}
				\BlankLine
				\fun{\mimic{$\A = \zug{\Sigma, Q, q_I, \delta, F}\colon$\normalfont commutative permutation \DFA, $p,q, s \in Q \setminus F$}}{
					\textit{Assumption:} $|\Sigma|$ \normalfont is fixed, let $\Sigma = \{\sigma_1, \sigma_2, \dots, \sigma_m\}$\;
					\ForEach
					(\ccc*[f]{\color{gray}possible since $|\Sigma|$ is fixed})
					{$1 \leq x_1 + \dots + x_{|\Sigma|} \leq |Q|$}{
						\lIf{$\delta(p, \sigma_1^{x_1}\sigma_2^{x_2}\dots\sigma_m^{x_m} ) = q$}{
							\Return{$\delta(s, \sigma_1^{x_1}\sigma_2^{x_2}\dots\sigma_m^{x_m} )$}
						}
					}
				}
				\BlankLine
				\fun{\mimic{$\A = \zug{\Sigma, Q, q_I, \delta, F}\colon$\normalfont commutative permutation \DFA, $p,q, s \in Q \setminus F$}}{
					\textit{Assumption:} \normalfont  this algorithm is allowed to use non-determinism\;
					$p' \affects p, \ell \affects 0$\;
					\While{$p' \neq q$ \and $\ell < |Q|$}{
						\guess $\sigma \in \Sigma$
						\ccc*[r]{\color{gray}iteratively contruct $w_{p,q}$ of length $\ell$}
						$p' \affects \delta(p', \sigma)$, $s \affects \delta(s, \sigma)$, $\ell \affects \ell +1$\;
					}
					\leIf
					(\ccc*[f]{\color{gray}check $q = \delta(p, w_{p,q})$})
					{$\ell = |Q|$}{
						\Return{\textit{error}}
					}{\Return{$s$}}
					%(\ccc*[f]{\color{gray}case $q = \delta(p, w_{p,q})$})
					%{\Return{$s$}}
				}
			}
			\caption{Deterministic and non-deterministic version of the algorithm solving the \Decomp problem for commutative permutation \DFAs.}
			\label{alg:decomp-Log}
		\end{algorithm}

		% \begin{lemma}
		% %	For arbitrary alphabets, the \textsc{Decomposition} problem for commutative permutation \DFAs is in \NLOGSPACE.
		% 	Given a commutative permutation \DFA $\A$ and a rejecting state $p$, we can determine the existence of a word covering $p$ in nondeterministic log-space. 
		% \end{lemma}
		\begin{proof}[Proof of Item \ref{item:NL}]
			We adapt the algorithm presented in the proof of Item~\ref{item:Logspace}, and use the terminology introduced there. We adapt it in the sense that we do not store the word $w_{p, q}$ but instead guess it again every time we want to apply it to some state $\delta(p, w_{p,q}^\lambda)$. Therefore, we store an extra copy of pointers to the states $p$ and $q$. 
			%	We guess $w_{p,q}$ by successively guessing the numbers $i_\ell$ of letters $\sigma_\ell$ and applying $\sigma_\ell^{i_\ell}$ to both states $p$ and the currently investigated state $\delta(p, w_{p,q}^j)$. 
			We guess $w_{{p,q}_\lambda}$ applied in iteration step $\lambda$ by successively guessing at most $|Q|$ letters $\sigma\in \Sigma$ and applying $\sigma$ to both states $p$ and the currently investigated state $\delta(p, w_{{p,q}_1}w_{{p,q}_2}\dots w_{{p,q}_{\lambda-1}})$. The counter on the number of guessed letters can be stored in $\log{|Q|}$ bits.
			We check that we actually reached $\delta(p, w_{{p,q}_1}^{\lambda})$ after guessing some $w_{{p, q}_\lambda}$ by checking whether the state $p$ is mapped to $q$. Note that the words $w_{{p, q}_i}$ that are guessed in different iteration steps $i$ of $\lambda$ might differ, but they are all equivalent in the sense that they impose the same transitions in the cycle induced by $w_{{p,q}_1}$ on $p$ since $\A$ is a commutative permutation \DFA.
			As the representation of $w_{p, q}$ was the only part using super-logarithmic space in the  algorithm described in Lemma~\ref{lemma:com-perm-decomp-LOGSPACE}, the claim follows. Both variants of the algorithm (deterministic and non-deterministic) are depicted in Algorithm~\ref{alg:decomp-Log}.
		\end{proof}
	\end{toappendix}
	
	\subsection{Proof of Theorem \ref{theorem:poly_family}}

	\begin{figure}
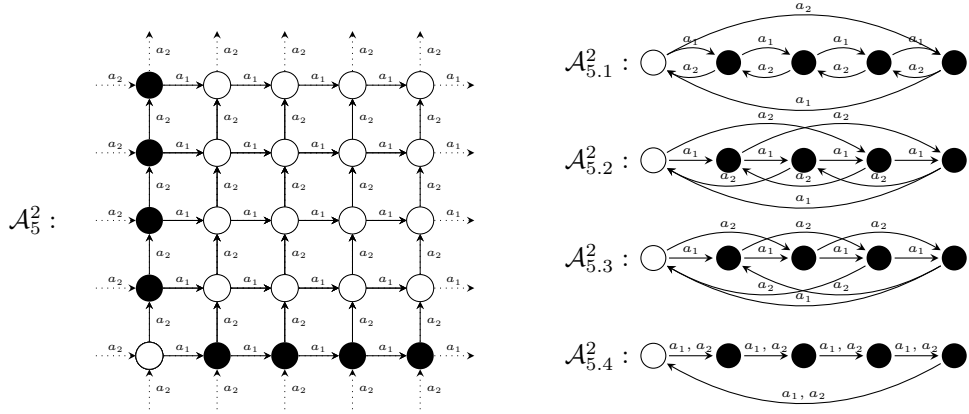
 %{r}{.45\linewidth}
		\centering
		\drawDecomposition
		\caption{
			The \DFA $\A_5^2$ recognising the language $L_5^2$, together with its decomposition
			into four non-trivial orbit-\DFAs.
			Final states are depicted in black.}\label{fig:decomposition}
	\end{figure}
	
	As a direct consequence of Lemma~\ref{lemma:Newword_dec=cover},
	the width of every commutative permutation \DFA $\A$
	is bounded by the number of rejecting states of $\A$, hence, it is smaller than $|\A|$.
	To conclude the proof of Theorem \ref{theorem:poly_family},
	for all $m,n \in \mathbb{N}$
	with $n$ prime,
	we define a \DFA $\A_n^m$ of size $n^m$ and width $(n-1)^{m-1}$
	on the alphabet $\Sigma = \{a_1,a_2, \ldots, a_m\}$.
	For all $\ell \in \N$,
	let $[\ell]$ denote the equivalence class
	of $\ell$ modulo $n$.
	Let $L_n^m \subseteq \Sigma^*$ be the language
	composed of the words $w$
	such that for at least one letter $a_i \in \Sigma$
	the number $\#_{a_i}(w)$ of $a_i$ in $w$
	is a multiple of $n$,
	and for at least one (other) letter $a_j \in \Sigma$,
	the number $\#_{a_j}(w)$ of $a_j$ in $w$
	is \emph{not} a multiple of $n$:
	\[
	L_n^m = \{w \in \Sigma^* \st
	[\#_{a_{i}}(w)] = [0] \textup{ and }
	[\#_{a_{j}}(w)] \neq [0] \textup{ for some }1 \leq i,j \leq m\}.
	\]
	The language $L_n^m$
	is recognised by a \DFA $\A_n^m$
	of size $n^m$ that keeps track
	of the value modulo $n$ of
	the number of each $a_i$
	already processed.
	The state space of $\A_n^m$
	is the direct product
	$(\mathbb{Z}/n\mathbb{Z})^m$
	of $m$ copies of the cyclic group 
	$\mathbb{Z}/n\mathbb{Z} = ([0],[1],\ldots,[n-1])$;
	the initial state is $([0],[0],\ldots,[0])$;
	the final states are the ones
	containing at least one component equal to $[0]$
	and one component distinct from $[0]$;
	and the transition function
	increments the $i^{\textup{th}}$
	component when an $a_i$ is read:
	$
	\delta(([j_1],[j_2],\ldots,[j_m]),a_i) = 
	([j_1],[j_2],\ldots,[j_{i-1}],[j_i+1],[j_{i+1}],\ldots,[j_m])$.
	Figure~\ref{fig:decomposition} illustrates the particular case $n=5$ and $m=2$.
	%@MAGICAPX
	
	To prove that the width of $\A_n^m$ is $(n-1)^{m-1}$,
	we first show that
	the $(n-1)^{m-1}$ words
	$\{a_1a_2^{\lambda_2} \ldots a_m^{\lambda_m}
	\st 1 \leq \lambda_i \leq n-1\}$ cover all the 
	rejecting states, thus by Lemma \ref{lemma:Newword_dec=cover}:
	%@MAGICAPX
	\begin{restatable}[\appendixProof]{proposition}{propComposite}\label{prop:composite}
		%\begin{restatable}{proposition}{propComposite}\label{prop:composite}
		The \DFA $\A_n^m$ is 
		$(n-1)^{m-1}$-factor composite.
	\end{restatable}

	\begin{toappendix}
		\propComposite*
		\begin{proof}
			For every $m-1$ tuple
			$\phi = (j_2,j_3,\ldots,j_m) \in \{1,2,\ldots,n-1\}^{m-1}$,
			let $u_{\phi}$ be the word
			\[
			u_{\phi} = a_1a_2^{j_2}a_3^{j_3}a_4^{j_4} \ldots a_m^{j_m} \in \Sigma^*.
			\]
			Note that there are $(n-1)^{m-1}$
			distinct words $u_\phi$.
			We show that every rejecting state of $\A_n^m$
			is covered by one of the $u_\phi$,
			which proves, by Lemma \ref{lemma:Newword_dec=cover}, that $\A_n^m$ is 
			$(n-1)^{m-1}$-composite.
			
			Let $q$ be a rejecting state of $\A_n^m$.
			Remember that the rejecting states of $\A_n^m$
			are precisely those for which either
			(\raisebox{.7pt}{\scalebox{.65}{$\blacklozenge$}})
			all the components are $[0]$,
			or
			(\raisebox{1pt}{\scalebox{.6}{$\bigstar$}})
			none of the component is $[0]$.
			We consider both possibilities.
			
			(\raisebox{.7pt}{\scalebox{.65}{$\blacklozenge$}})
			If $q = ([0],[0],\ldots,[0])$
			we show that every
			$u_{\phi} = a_1a_2^{j_2}a_3^{j_3} \ldots a_m^{j_m}$ covers $q$:
			for all $\lambda \in \mathbb{N}$,
			\[
			\begin{array}{lll}
				\delta(q,u_\phi^\lambda)
				& = &
				\delta(([0],[0],\ldots,[0]),
				(a_1a_2^{j_2}a_3^{j_3}a_4^{j_4} \ldots a_m^{j_m})^\lambda)\\
				& = &
				([\lambda],[\lambda j_2],[\lambda j_3],[\lambda j_4],\ldots,[\lambda j_m]).
			\end{array}
			\]
			Therefore, either $[\lambda] = [0]$
			and all the components of $\delta(q,u_\phi^\lambda)$
			are $[0]$,
			or $[\lambda] \neq [0]$
			and none of the components of $\delta(q,u_\phi^\lambda)$
			are $[0]$ (since $n$ is prime and $1<j_i<n-1$).
			In both cases, $\delta(q,u_\phi^\lambda)$ is rejecting.
			This proves that $u_\phi$ covers $q$.
			
			(\raisebox{1pt}{\scalebox{.6}{$\bigstar$}})
			If $q = ([k_1],[k_2],\ldots,[k_m])$
			such that none of the $[k_i]$ is equal to $[0]$,
			we build a specific $u_\phi$ that covers $q$.
			Since $[k_1] \neq 0$ and $n$ is prime,
			there exists $\mu \in \N$
			satisfying $[\mu \cdot k_1] = [1]$.
			Note that this implies that $[\mu] \neq [0]$,
			hence for every $2 \leq i \leq m$
			we get that $[\mu k_i]=[j_i]$ for some $1 \leq j_i \leq n-1$.
			Let $\phi = (j_2,j_3,\ldots,j_m)$.
			Then for every $\lambda \in \mathbb{N}$
			\[
			\begin{array}{lll}
				\delta(q,u_\phi^\lambda)
				& = &
				\delta(([k_1],[k_2],\ldots,[k_m]),
				(a_1a_2^{j_2}a_3^{j_3}a_4^{j_4} \ldots a_m^{j_m})^\lambda)\\
				& = &
				([k_1 + \lambda],[k_2 + \lambda j_2],[k_3 + \lambda j_3],[k_4 + \lambda j_4],\ldots,[k_m + \lambda k_m])\\
				& = &
				([k_1 + \lambda \mu k_1],[k_2 + \lambda\mu k_2],[k_3 + \lambda\mu k_3],[k_4 + \lambda\mu k_4],\ldots,[k_m + \lambda\mu k_m])\\
				& = &
				([(1+\lambda \mu)k_1],[(1+\lambda \mu)k_2],[(1+\lambda \mu)k_3],[(1+\lambda \mu)k_4],\ldots,[(1+\lambda \mu)k_m]).
			\end{array}
			\]
			Remember that, by supposition,
			$[k_i] \neq [0]$ for all $1 \leq i \leq m$.
			Therefore, either $[\lambda \mu + 1] = [0]$
			and all the components of $\delta(q,u_\phi^\lambda)$
			are $[0]$,
			or $[\lambda \mu + 1] \neq [0]$
			and none of the components of $\delta(q,u_\phi^\lambda)$
			are $[0]$ (since $n$ is prime).
			In both cases, $\delta(q,u_\phi^\lambda)$ is rejecting.
			This proves that $u_\phi$ covers $q$.
		\end{proof}
	\end{toappendix}
	
	% \begin{proposition}\label{prop:not_composite}
	%  $u$ covers $([i],[j])$ if and only if
	%  $\#_a(u) \cdot j \equiv \#_b(u) \cdot i$ (mod $n$).
	% \end{proposition}
	
	% \begin{proof}
	%     Suppose that
	%     $\#_a(u) \cdot j \not \equiv \#_b(u) \cdot i$ (mod $n$)

	%     Suppose that
	%     $\#_a(u) \cdot j \equiv \#_b(u) \cdot i$ (mod $n$).
	%     Then for every $\lambda \in \mathbb{N}$,
	%     \[
	%     \delta(([i],[j]),u^{\lambda}) =
	%     \delta(([i + \lambda \#_a(u)],[j + \lambda \#_b(u)]),u^{\lambda})
	%     \]
	%     \[
	%     [i + \lambda \#_a(u)]=[0]
	%     \Leftrightarrow
	%     i \equiv \lambda \#_a(u)
	%     \Leftrightarrow
	%     i \equiv \lambda \#_a(u)
	
	%     \]
	
	% \end{proof}

	Then,
	%\nicolas{space}
	we prove that
	there exist no word that covers
	two states among the $(n-1)^{m-1}$
	rejecting states
	$\{([1],[k_2],[k_3],\ldots,[k_m]) \st 1 \leq k_i \leq m-1\}$.
	Therefore, we need at least $(n-1)^{m-1}$ words
	to cover all of the states,
	thus by Lemma \ref{lemma:Newword_dec=cover}:
	%@MAGICAPX
	\begin{restatable}[\appendixProof]{proposition}{propNotComposite}\label{prop:not_composite}
		%\begin{restatable}{proposition}{propNotComposite}\label{prop:not_composite}
		The \DFA $\A_n^m$ is not  $((n-1)^{m-1}-1)$-factor composite.
	\end{restatable}
	
	\begin{toappendix}
		\propNotComposite*
		\begin{proof}
			For every $m-1$ tuple
			$\phi = (k_2,k_3,\ldots,k_m) \in \{1,2,\ldots,n-1\}^{m-1}$,
			let $q_\phi$ denote the rejecting state
			$([1],[k_2],[k_3],\ldots,[k_m])$
			of $\A_n^m$.
			Note that there are $(n-1)^{m-1}$
			distinct $q_\phi$.
			We show that there exists no word
			that covers two different $q_\phi$,
			which proves, by Lemma \ref{lemma:Newword_dec=cover},
			that $\A_n^m$ is not
			$((n-1)^{m-1}-1)$-composite.
			
			Let
			$\phi = (k_2,k_3,\ldots,k_m),\psi = (\ell_2,\ell_3,\ldots,\ell_m) \in \{1,2,\ldots,n-1\}^{m-1}$,
			and let $u \in \Sigma^*$
			be a word that covers both $q_{\phi}$ and $q_{\psi}$.
			We show that this implies $\phi = \psi$.
			Since $\A_{n}^m$
			is commutative,
			we can suppose
			without loss of generality that
			$u = a_1^{j_1}a_2^{j_2} \ldots a_m^{j_m}$.
			Let $\lambda \in \mathbb{N}$
			satisfying $[\lambda j_1] = [-1]$.
			Then 
			\[
			\begin{array}{lll}
				\delta(q_\phi,u^\lambda)
				& = &
				([0],[k_2 + \lambda j_2],[k_3 + \lambda j_3],[k_4 + \lambda j_4],\ldots,[k_m + \lambda j_m])\\
				\delta(q_\psi,u^\lambda)
				& = &
				([0],[\ell_2 + \lambda j_2],[\ell_3 + \lambda j_3],[\ell_4 + \lambda j_4],\ldots,[\ell_m + \lambda j_m]).
			\end{array}
			\]
			Since $u$ covers both $q_\phi$
			and $q_\psi$ by supposition,
			both $\delta(q_\phi,u^\lambda)$
			and $\delta(q_\psi,u^\lambda)$
			are rejecting states of $\A_n^m$.
			Since the first component
			of both of these states is $[0]$,
			this implies that
			\emph{all} of their components are $[0]$.
			In other words,
			for every $2 \leq i \leq m$ we get
			$[k_i + \lambda j_i] = [0] = [\ell_i + \lambda j_i]$,
			hence $k_i = \ell_i$.
			This proves that $\phi = \psi$.
		\end{proof}
	\end{toappendix}

	\section{Bounded Decomposition}\label{sec:bounded_decomposition}
	
	We finally study
	the \boundDecomp
	problem: Given a \DFA $\A$ and an integer $k \in \N$ encoded in unary,
	can we determine whether $\A$
	is decomposable into $k$ factors?
	For the general setting, we show
	that the problem is in \PSPACE:
	it can be solved by non-deterministically 
	guessing $k$ factors,
	and checking that they form a decomposition.
	
	%@MAGICAPX
	\begin{restatable}[\appendixProof]{theorem}{BoundedDec}
		%\begin{restatable}{theorem}{BoundedDec}
		\label{thm:bounded_decomposition}
		The \boundDecomp problem is in \PSPACE.
		%	It is solvable in $\Sigma^P_2$ if $\A$ has a singleton alphabet.
	\end{restatable}
	\begin{toappendix}
		\BoundedDec*
		\begin{proof}
			%	\todo[inline,color=blue!20]{Isn't it enough to check whether $L(\A)$ is included into each $L(\A_i)$, and whether
			%		$\overline{L(\A)} \cap \bigcap_{i=1}^kL(\A_i) = \varnothing$?
			%		I don't think that we need to build the product,
			%		even implicitely.\\
			%		\textcolor{magenta}{True, should I adapt it?}}
			
			For a language $L \subseteq \Sigma^*$ we denote with $\overline{L}$ the complement $\Sigma^* \setminus L$ of $L$.
			Let $\A=\zug{\Sigma,Q_\A,{q_I}_\A,\delta_\A,F_\A}$ be a \DFA and let $k \in \N$ be encoded in unary.
			We non-deterministically guess $n \leq k$ \DFAs $\A_1, \A_2, \dots, \A_n$ with $\A_i = \zug{\Sigma,Q_{\A_i},q_{I_{\A_i}},\delta_{\A_i},F_{\A_i}}$ for $1 \leq i \leq n$, such that $|\A_i| < |\A|$. We implicitly build the product \DFA $\Pi_1^n \A_i = \A_1 \times \A_2 \times \dots \A_n$ over the state space $Q_{\A_1} \times Q_{\A_2} \times \dots \times Q_{\A_n}$ with the start state $(q_{I_{\A_1}}, q_{I_{\A_2}}, \dots, q_{I_{\A_n}})$ and set of final states $F_{\A_1} \times F_{\A_2} \times \dots \times F_{\A_n}$, where in the $i$'th component the run of the \DFA $\A_i$ on the input is simulated. We do not build this \DFA explicitly as it is of exponential size in $|Q_\A|$.
			Note that $\Pi_1^n \A_i$ accepts $\bigcap_{1 \leq i \leq n}L(\A_i)$. 
			%	
			%	Let us denote this \DFA with $\mathcal{C}(\Pi_1^n \A_i)$. 
			In order to prove whether $L(\A) = L(\Pi_1^n \A_i)$ it is sufficient to verify that (1) $L(\A) \cap \overline{L( \Pi_1^n \A_i)} = \varnothing$ and (2) $\overline{L(\A)} \cap L(\Pi_1^n \A_i) = \varnothing$. 
			%	For a language $L$ given by a \DFA $\B$ we know that $L \neq \varnothing$ if and only if some final state of $\B$ is reachable from the initial state of $\B$. As reachability for directed graphs can be decided in non-deterministic log-space 
			As $\Pi_1^n \A_i$ is a \DFA, we can obtain a \DFA for the complementary language $\overline{L(\Pi_1^n \A_i)}$ by complementing on the set of final states. 
			We can test the complementary statement of both (1) and (2) by letter-wise guessing a word in the intersection and applying its map on the initial state of both \DFAs. As we only need to store the active state of both \DFAs, this can be done in \NPSPACE. As \NPSPACE is closed under complement and is equal to \PSPACE, the claim follows.
		\end{proof} 
	\end{toappendix}
	
	For
	%\nicolas{space}
	commutative permutation \DFAs,
	we obtain a better algorithm through
	the use of the results obtained in the
	previous sections,
	and we show a matching hardness result. 
	
	\begin{theorem}\label{thm:bounded_dec_NP}
		The \boundDecomp
		problem for commutative permutation \DFAs
		is \NP-complete.
	\end{theorem}
	
	\noindent
	Both parts of the proof of Theorem \ref{thm:bounded_dec_NP}
	are based on Lemma \ref{lemma:Newword_dec=cover}:
	a commutative permutation \DFA
	is $k$-factor composite
	if and only if there exist
	$k$ words covering all of
	its rejecting states.
	We prove the two following results:
	\begin{itemize}
		\item
		Bounded compositionnality
		is decidable in \NP, as it is sufficient to
		non-deterministically guess
		a set of $k$ words,
		and check whether they cover all rejecting states (Lemma \ref{lemma:bounded_dec_NP});
		\item
		The \NP-hardness is obtained by reducing
		the \textsc{Hitting Set} problem,
		a well known \NP-complete decision problem.
		We show that searching for $k$ words that cover
		the rejecting states of a \DFA is as complicated as
		searching for a hitting set of size $k$
		(Lemma \ref{lemma:bounded_dec_hard}).
	\end{itemize}
	
	\noindent
	We finally give a \LOGSPACE algorithm based on known results for \DFAs on unary alphabets~\cite{JKM20}.
	
	%@MAGICAPX
	\begin{restatable}[\appendixProof]{theorem}{BoundedDecUnary}
		%\begin{restatable}{theorem}{BoundedDecUnary}
		\label{thm:bounded_decomposition_unary}
		The \boundDecomp
		problem for unary \DFAs
		is in \LOGSPACE.
	\end{restatable}
	\begin{proof}[Sketch]
		Recall that a unary \DFA $\A=\zug{\{a\},Q,q_I,\delta,F}$ consists of a chain of states leading into one cycle of states. The case where the chain is non-empty is considered in Lemmas~8 and 10 of~\cite{JKM20}. We prove that the criteria of these lemmas can be checked in \LOGSPACE. If the chain of $\A$ is empty, then $\A$ is actually a commutative permutation \DFA. In this case, by Proposition~\ref{prop:disjoint} for every word $u=a^i \in \{a\}^*$, the orbit of the set $\{\delta(q_I, u^\lambda) \st \lambda \in \mathbb{N}\}$ is a partition $\rho$ on $Q$, and every set in $\rho$ has the same size $s_\rho$.
		Both $s_\rho$ and $|\rho|$ divide $|Q|$. 
		For $u=a^i$ where $i$ and $|Q|$ are co-prime, the induced orbit \DFA has a single state and thus cannot be a factor of $\A$. Further, if $i_1 < |Q|$ divides $i_2 < |Q|$, then all states covered by $a^{i_1}$ are also covered by $a^{i_2}$.
		Hence, w.l.o.g., we only consider words of the form $a^i$ where $i$ is a \emph{maximal divisor} of $|Q|$ in order to generate orbit-\DFAs of $\A$ that are candidates for the decomposition.
		Now, let $p_1^{j_1} \cdot p_2^{j_2} \cdot\ldots \cdot p_m^{j_m} = |Q|$ be the prime factor decomposition of $|Q|$.
		By Lemma~\ref{lemma:Newword_dec=cover} we have that $\A$ is $k$-factor composite if and only if a selection of $k$ words from the set $\mathcal{W} = \{a^{|Q|/ p_i} \st 1 \leq i \leq m \}$ cover all the rejecting states of $\A$. As $|\mathcal{W}| = m$ is logarithmic in $|Q|$, we can iterate over all sets in $2^\mathcal{W}$ of size at most $k$ in \LOGSPACE using a binary string indicating the characteristic function. By Lemma~\ref{lemma:com-perm-decomp-LOGSPACE}, checking whether a state $q\in Q$ is covered by the current collection of $k$ words can also be done in \LOGSPACE.
	\end{proof}
	\begin{toappendix}
		\BoundedDecUnary*
		
		In order to prove the theorem, we first show that the result holds for unary permutation \DFAs,
		and then we show how this extends to the general setting.
		
		\begin{lemma}
			\label{thm:unary-perm-k-factor-logspace}
			The \boundDecomp
			problem for permutation unary \DFAs
			is in \LOGSPACE.
		\end{lemma}
		\begin{proof}
			Let $\A=\zug{\{a\},Q,q_I,\delta,F}$ be a trim permutation \DFA. Since the alphabet of $\A$ is unary, $\A$ is also commutative. Hence, by Proposition~\ref{prop:disjoint} for every word $u=a^i \in \{a\}^*$ the orbit of the set $\{\delta(q_I, u^\lambda) \st \lambda \in \mathbb{N}\}$ is a partition $\rho$ on $Q$. Since $\A$ is a permutation \DFA, every word induces a permutation of $Q$ and therefore, every set in the partition $\rho$ has the same size $s_\rho$. Hence, both $s_\rho$ and $|\rho|$ divide $|Q|$. Further, note that for every integer $i>1$ there is at most one partition $\rho$ of $Q$ of size $|\rho|=i$
			that is consistent with the transition relation of $\A$,
			and  this partition (if existent) corresponds to the orbit-\DFA generated by the word $a^{|Q|/i}$ where $i$ evenly divides $|Q|$. 
			For $i=1$, $\{\delta(q_I, u^\lambda) \st \lambda \in \mathbb{N}\} = Q$ corresponds to the trivial orbit-\DFA which has only one state. This trivial partition is generated by every word $a^{i}$ where $i$ and $|Q|$ are co-prime.
			As trivial orbit-\DFAs do not contribute to a decomposition, it is sufficient to consider only words $a^i$ where $i$ is a divisor of $|Q|$ to generate all orbit-\DFAs of $\A$ that needs to be considered in order to obtain a $k$-factor decomposition of $\A$. Further, note that for two integers $i_1$ and $i_2$ it holds that if $i_1$ divides $i_2$, then all states covered by $a^{i_1}$ are also covered by $a^{i_2}$.
			Therefore,
			while looking for covering words
			it is sufficient to look through the
			\emph{maximal} divisors of $|Q|$, that is,
			the divisors that do not divide other divisors.
			
			Now, let $p_1^{j_1} \cdot p_2^{j_2} \cdot\ldots \cdot p_m^{j_m} = |Q|$ be the prime factor decomposition of $|Q|$, i.e., $p_i$ are prime numbers for $1 \leq i \leq m$.
			By the discussion above, and Lemma~\ref{lemma:Newword_dec=cover} we have that $\A$ is $k$-factor composite if and only if a selection of $k$ words from the set $\mathcal{W} = \{a^{|Q|/ p_i} \st 1 \leq i \leq m \}$ cover all the rejecting states of $\A$. 
			Note that $|\mathcal{W}|=m$ is logarithmic in $|Q|$ and hence, the size of the power set $2^{\mathcal{W}}$ of $\mathcal{W}$ is linear in $|Q|$.
			Further, we can represent a set in $2^\mathcal{W}$ as a binary string (with a 1 at position $i$ iff the $i$th element is in the set represented by the string) of size $m$ and iterate through $2^\mathcal{W}$ in logarithmic space. 
			We can now check whether $\A$ is $k$-factor composite by iterating through $2^\mathcal{W}$ and testing whether for one set of words, each rejecting state of $\A$ is covered by one of the words in the considered set.
			How to verify that a rejecting state is covered by a word in logarithmic space is discussed in the proof of Lemma~\ref{lemma:com-perm-decomp-LOGSPACE}.
			Note that since $|Q|$ is given in unary, we can compute a prime divisor $p_i$ of $|Q|$ in logarithmic space when needed and only need to store the currently considered word $a^{|Q|/p_i}$ in binary.
			The described \LOGSPACE-algorithm is summarized in the first case of Algorithm~\ref{alg:unary-k-factor-Log}.
		\end{proof}
		
		General unary \DFAs consist of a cycle and a potentially empty chain of states from the initial state into the cycle. If this chain is empty, the \DFA is actually a permutation \DFA. If the tail is non-empty, then the \DFA $\A$ is composite if and only if $\A$ is $2$-composite, or not minimal (and hence $1$-composite) due to Lemma 8-10 in~\cite{JKM20}. The criteria in~\cite[Lemma~8]{JKM20} and \cite[Lemma~9]{JKM20} can obviously be checked in \LOGSPACE. The remaining criteria of \cite[Lemma~10]{JKM20} considers unary \DFAs $\A$ where the two preimages of the state in the cycle, connecting the cycle with the chain, are separated by the set of final states. If the preimage from the chain is accepting and the preimage $q_c$ from the cycle is rejecting, then \cite[Lemma~10]{JKM20} states that $\A$ is composite if and only it is 2-composite if and only if the state $q_c$ is covered by some word $w$ in the commutative permutation sub-automaton consisting of the cycle only. The latter case can be checked in logarithmic space as a consequence of Lemma~\ref{thm:unary-perm-k-factor-logspace} yielding in summary a proof of Theorem~\ref{thm:bounded_decomposition_unary}. The complete algorithm solving the \boundDecomp problem for unary \DFAs in \LOGSPACE is depicted in Algorithm~\ref{alg:unary-k-factor-Log}.

		% ++++++++++++ IMPORT ALGO ++++++++++++
		%\begin{figure}[h]
		%\centering
		%\includegraphics{algo_kdec_unary.png}
		%\stepcounter{theorem}
		%\end{figure}
		% ++++++++++++ ++++++++++++ +++++++++++
		\begin{algorithm}[]
			\showalgo{
				\DontPrintSemicolon
				\SetKwFunction{main}{\normalfont isBoundedComposite}
				\SetKwFunction{cover}{\normalfont cover}
				\SetKw{true}{\normalfont True}
				\SetKw{false}{\normalfont False}
				% %
				\SetKw{guess}{guess}
				\SetKw{neg}{not}
				\SetKw{compute}{compute}
				% %
				\SetKwFunction{coverBySet}{\normalfont coverBySet}
				\SetKwFunction{testWordComb}{\normalfont testWordCombination}
				\SetKwFunction{coverLog}{\normalfont coverLog}
				\SetKwFunction{bin}{\normalfont bin}
				\SetKw{int}{int}
				\SetKw{binString}{binaryString}
				\SetKw{break}{break}
				\SetKw{call}{call}
				\SetKw{wordCombination}{\normalfont wordCombination}
				\SetKw{badSet}{\normalfont badSetFlag}
				% %
				\SetKwProg{fun}{Function}{}{end}	
				\SetKwComment{ccc}{\color{gray}/* }{\color{gray}\ */}
				
				\fun{\main{$\A = \zug{\{a\}, Q, q_I, \delta, F}\colon$\normalfont unary \DFA, integer $k$}}{	
					
					\eIf{$\A$ is permutation \DFA}{
						
						\ForEach(\ccc*[f]{\color{gray}\wordCombination represents current set in $2^\mathcal{W}$}){\binString \wordCombination $\in \{0, 1\}^{\log |Q|}$ with $\leq k$ ones}
						{
							
							\lIf{\testWordComb{$\A$, \wordCombination}}{\Return{\true}}\ccc*[f]{\color{gray}Set of words covering all rejecting states found}
							
						}
						\Return{\false}\ccc*[r]{\color{gray}No covering set found}
					}{\call~\cite[Algorithm~1]{JKM20}}
					
				}
				
				\fun{\testWordComb{$\A = \zug{\{a\}, Q, q_I, \delta, F}\colon$\normalfont unary \DFA, $\wordCombination\colon\binString$}}{	
					\ForEach{$q\in Q\setminus F$}{
						
						\lIf(\ccc*[f]{\color{gray}Found state not covered by current set}){\neg \cover($\A, q, \wordCombination$)}{\Return{\false}}
						
					}
					\Return{\true}		
				}
				
				\fun{\coverBySet{$\A = \zug{\{a\}, Q, q_I, \delta, F}\colon$\normalfont unary \DFA, $q\in Q\setminus F$, $\wordCombination\colon\binString$}}{	
					\ForEach(\ccc*[f]{\color{gray}Go through all $\leq k$ words in the set and test if $q$ is covered}){\int $i$ with $\wordCombination[i] \testeq 1$}{

						\compute $p_1\affects i$'th prime divisor of $|Q|$
						
						\lIf(\ccc*[f]{\color{gray}Function \cover from Algorithm~\ref{alg:decomp-Log}}){\cover{$\A, q, \delta(q, a^{|Q|/p_i})$}}
						{\Return{\true}}	
						
					}
					\Return{\false}
				}
				
			}
			\caption{\LOGSPACE-algorithm solving the \boundDecomp problem for unary \DFAs.}
			\label{alg:unary-k-factor-Log}
		\end{algorithm}
	\end{toappendix}
	
	\subsection{Proof of Theorem \ref{thm:bounded_dec_NP}}
	%\myparagraph{Proof of Theorem \ref{thm:bounded_dec_NP}} The rest of this section is denoted to the proof of Theorem~\ref{thm:bounded_dec_NP}.
	%We begin with a straightforward \NP algorithm for the \boundDecomp problem.
	%For the \boundDecomp problem of commutative permutation \DFAs,
	By Lemma~\ref{lemma:Newword_dec=cover}, a commutative permutation \DFA $\A$ is $k$-factor composite 
	if and only if its rejecting states can be covered by $k$ words.
	As we can suppose that covering words have size linear in~$|\A|$
	%
	%Since we can suppose that the size of the covering words is linear with respect to $|\A|$ 
	(see proof of Lemma~\ref{lemma:com-perm-decomp-LOGSPACE}),
	the \boundDecomp problem is decidable in \NP:
	we guess a set of $k$ covering words and check in polynomial time
	if they cover all rejecting states. 
	
	%@MAGICAPX
	\begin{restatable}[\appendixProof]{lemma}{BoundedDecComPermNP}
		%\begin{restatable}{lemma}{BoundedDecComPermNP}
		\label{lemma:bounded_dec_NP}
		The \boundDecomp
		problem for commutative permutation \DFAs
		is in \NP.
	\end{restatable}
	\begin{toappendix}
		\BoundedDecComPermNP*
		\begin{proof}
			Let $\A=\zug{\Sigma,Q,q_I,\delta,F}$ be a commutative permutation \DFA.
			By Lemma~\ref{lemma:Newword_dec=cover} we have that $\A$ is $k$-factor composite
			if and only if there are $k$ words that, together, cover all rejecting states of $\A$. 
			Recall that a word $w$ covers a state $q$ if $\delta(q, w) \neq q$ and for every $\lambda\in \N$, the state $\delta(q, u^\lambda)$ is rejecting.
			Since $\A$ is a commutative permutation \DFA, for each word $w \in \Sigma^*$ with $|w|>|Q|$ there exists a word $u \in \Sigma^*$ with $|u|<|Q|$ that induces the same mapping as $w$, in particular, $\delta(q, w^\lambda) = \delta(q, u^\lambda)$ for all $q \in Q$, $\lambda \in \N$.
			Hence, it is sufficient to test whether the rejecting states of $\A$ can be covered by $k$ words of length up to $|Q|$. As we can guess these words $u_i$ in polynomial time and check whether they cover all rejecting states $q$ by computing the sets $\{\delta(q, u^\lambda) \st \lambda \leq |Q|\}$ in polynomial time, the claim follows. The procedure is summarized in Algorithm~\ref{alg:com-perm-k-factor-NP}.
		\end{proof}
	\end{toappendix}
	
	% ++++++++++++ IMPORT ALGO ++++++++++++
	%\begin{figure}[h]
	%\centering
	%\includegraphics{algo_kdec.png}
	%\stepcounter{theorem}
	%\end{figure}
	% ++++++++++++ ++++++++++++ +++++++++++
	\begin{toappendix}
		\begin{algorithm}[]
			\showalgo{
				\DontPrintSemicolon
				\SetKwFunction{main}{\normalfont isBoundedComposite}
				\SetKwFunction{cover}{\normalfont cover}
				\SetKwProg{fun}{Function}{}{end}
				\SetKw{true}{\normalfont True}
				\SetKw{false}{\normalfont False}
				\SetKw{neg}{not}
				\SetKw{guess}{guess}
				\SetKw{compute}{compute}
				\SetKwComment{ccc}{\color{gray}/* }{\color{gray}\ */}
				
				\fun{\main{\normalfont commutative permutation \DFA $\A$, integer $k$}}{

					\guess{$\mathcal{W} := \{w_i \in \Sigma^{\leq |Q|} \st i \leq k\}$}
					
					\ForEach{$p \in Q \setminus F$}{
						%				\coverExists \affects \false\;
						%				\ForEach{$w_i \in \mathcal{W}$}{
						
						\lIf(\ccc*[f]{\color{gray}Some $p$ not covered?}){\neg \cover{$\A, p, \mathcal{W}$}}{
							\Return{\false}}
					}
					%				\lIf{\neg \coverExists}{\Return{\false}\tcc*[f]{No $w_i$ coverse $p$}
					%				}
					%			}
					
					\Return{\true}\ccc*[r]{\color{gray}all $p$ are covered}
					
					%			\lIf{$s_{k-1} \in F \Leftrightarrow q_{\ell-1} \in F$}{
					%				\Return{false}
					%				\tcc*[f]{$\A$ is not minimal}
					%			}
					%		
					%		
					%			\funbool{\isCompDFA{\normalfont commutative permutation \DFA $\A$, \int $k$}}{	
					%			
					%			
					%			\guess{$\mathcal{W} := \{w_i \in \Sigma^{\leq |Q|} \st i \leq k\}$}
					%			
					%			\ForEach{$p \in Q \setminus F$}{
					%			
					%				
					%			\lIf{\neg\cover{$\A, q, w_i$}}{
					%				\Return{\false}
					%				\tcc*[f]{Does $w_i$ cover $q$?}}
					%			}
					%		
					%			\Return{\true}
					%		
					%%			\lIf{$s_{k-1} \in F \Leftrightarrow q_{\ell-1} \in F$}{
					%%				\Return{false}
					%%				\tcc*[f]{$\A$ is not minimal}
					%%			}
					%%		
					%%			\lIf{$s_{k-1} \in R$ \and $q_{\ell-1} \in F$}{
					%%				\Return{$\ell \testeq 1$}
					%%				\tcc*[f]{by the $k > 0$ case}
					%%			}
					%%			
					%%			\uIf{$k \testeq 0$}{
					%%				\Return{\hasBadQuotient{$\A$}}
					%%			}
					%%		
					%%			\If{$s_{k-1} \in F$ \and $q_{\ell-1} \in R$}{
					%%				\Return{\hasBadQuotient{$\A$}}
					%%			}
					%		
					%		}
					%	
					%		\funbool{\cover{\normalfont commutative permutation \DFA $\A$, state $q$, word $w_i$}}{
					%			\ForEach{$w_i \in \mathcal{W}$}{
					%			
					%				\compute $Q_{q, w_i}:=\{\delta(q, w_i^\lambda)\mid \lambda \leq |Q|\}$
					%				
					%				\lIf{$Q_{q, w_i} \cap F = \emptyset$}{\Return{\true}}
					%			}
					
				}
				
				\fun{\cover{\normalfont commutative permutation \DFA $\A$, state $p$, set of words $\mathcal{W}$}}{
					\ForEach{$w_i \in \mathcal{W}$}{
						
						\compute $Q_{q, w_i}:=\{\delta(q, w_i^\lambda)\mid \lambda \leq |Q|\}$
						
						\lIf{$Q_{q, w_i} \cap F = \varnothing$}{\Return{\true}}
					}
					\Return{\false}
				}		
			}
			\caption{\NP-algorithm solving the \boundDecomp problem for commutative permutation \DFAs.}
			\label{alg:com-perm-k-factor-NP}
		\end{algorithm}
	\end{toappendix}
	\noindent
	We show that the problem is \NP-hard
	by a reduction from  the \textsc{Hitting Set} problem.
	\begin{lemma}\label{lemma:bounded_dec_hard}
		The \boundDecomp problem
		is \NP-hard for commutative permutation
		\DFAs.
	\end{lemma}
	\begin{proof}
		The proof goes by a reduction from the \textsc{Hitting Set} problem (\HIT for short), known to be \NP-complete \cite{10.5555/578533}.
		The \HIT problem asks, given a finite set $S = \{1, 2, \dots, n\} \subseteq \N$, a finite collection of subsets $\mathcal{F}= \{C_1, C_2, \dots, C_m\} \subseteq 2^S$, and an integer $k \in \N$, whether there is a subset $X \subseteq S$ with $|X| \leq k$ and $X \cap C_i \neq \varnothing$ for all $1 \leq i \leq m$.
		%Let us consider the instance of \HIT with $S = \{1, 2, \dots, n\}$,  $\mathcal{F} = \{C_1, C_2\dots, C_m\}$ and $k \in \N$.
		We describe how to construct a \DFA  $\A=\zug{\Sigma,Q,q_I,\delta,F}$ that is $(k+1)$-factor composite if and only if the \HIT instance $\zug{S, \mathcal{F}, k}$ has a solution.
		%
		%\textsc{Hitting Set Problem} (\HIT for short)\red{REF!!}
		%
		%Given: A set $S = \{1, 2, \dots, n\} \subseteq \N$,
		%a finite collection of sets $\mathcal{F}= \{C_1, C_2, \dots, C_m\}$ with $C_i \subseteq S$ for $1\leq i \leq m$, and an integer $k$ with $1 \leq k \leq n$.
		%\hangindent=0.8cm
		%
		%Question: Is there a subset $X \subseteq S$ with $|X| \leq k$ and $X \cap F_i \neq \varnothing$ for all $1 \leq i \leq m$?
		%
		\begin{figure}
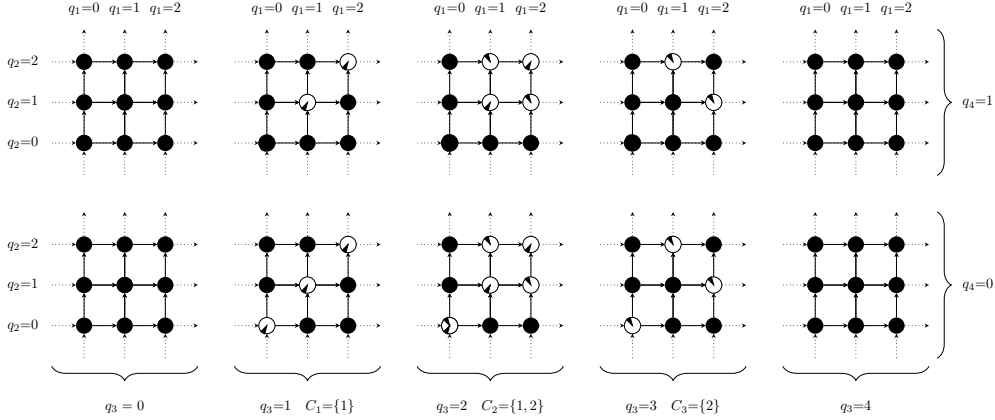

			\centering
			\scalebox{.54}{\drawHard}
			\caption{\DFA representing the instance of \HIT with $S = \{1, 2\}$ and $\mathcal{F} = \{\{1\}, \{1,2\}, \{2\}\}$ using $\p = 3$ and $\q = 5$. Accepting states are filled black while rejecting states are sectored.}\label{fig:hard:example}
		\end{figure}

		\boldtitle{Automaton construction}
		To be constructed, the automaton $\A$ requires $\p, \q$ defined as the smallest prime numbers that fulfill $n < \p$ and $m < \q$ and $2 < \p < \q$.
		By Bertrand's postulate~\cite{doi:10.4169/amer.math.monthly.120.07.650}, $\p$ and $\q$ have a value polynomial in $m+n$.
		The state space of $\A$ is defined as $Q = \{0, 1, \dots, \p-1 \} \times \{0, 1, \dots, \p-1 \} \times \{0, 1, \dots, \q-1 \} \times \{0, 1\}$ with $q_I = (0, 0, 0, 0)$ as initial state.
		Let us define the subset of states $Q_\bot = \{(q_1, q_2, q_3, q_4) \in Q \st q_4 = 0\}$ to encode instances of \HIT and the subset $Q_\top = \{(q_1, q_2, q_3, q_4) \in Q \st q_4 = 1\}$ which is a copy of $Q_\bot$ with minor changes.
		The example in Figure~\ref{fig:hard:example} gives some intuition on the construction of $\A$.
		The \DFA $\A$ is defined over the alphabet $\Sigma = \{a, b, c, d\}$ with the transition function defined for each state $q = (q_1, q_2, q_3, q_4)$ by $\delta(q, a) = (q_1+1 \mod \p,q_2,q_3,q_4)$, $\delta(q, b) = (q_1,q_2+1 \mod \p,q_3,q_4)$, $\delta(q, c) = (q_1,q_2,q_3+1 \mod \q,q_4)$ and $\delta(q, d) = (q_1,q_2,q_3,q_4+1\mod 2)$.
		Note that, $\A$ can be seen as a product of four  prime finite fields.
		In particular, for every $q_3 \in \{0, \dots, \q-1\}$ the subset of states $\{(x,y,q_3,0) \in Q_\bot \st 0 \leq x,y \leq \p-1\}$ can be seen as the direct product of two copies of the field of order $\p$ (a.k.a.\ $\mathbb{F}_{\p}$), thus inheriting the structure of a $\mathbb{F}_{\p}$-vector space of origin $(0,0,q_3,0)$.
		We use these $\q$ disjoint vector spaces to represent the collections of $\mathcal{F}$ thanks to the acceptance of states.
		%The elements of the collection $\mathcal{F}$ are represented by $\q > m$ disjoint vector planes.
		More precisely, each collection $C_i\in \mathcal{F}$ is encoded through the vector space $\{(x, y, i, 0) \in Q_\bot \st 1 \leq i \leq m\}$ and each $v \in C_i$ is encoded by the non-acceptance of all states belonging to the line $\{(x, y, i, 0) \in Q_\bot \st y = vx \mod \p\}$.
		In Figure~\ref{fig:hard:example}, each $C_i$ is presented by an instance of $\mathbb{F}_3\times\mathbb{F}_3$ and each $v \in \C_i$ is depicted by rejecting states with the same emphasized sector.
		Since $\q > m$, there are extra vector spaces for which all states are accepting i.e.\ $\{ (q_1, q_2, q_3, 0) \in Q_\bot \st q_3 \notin \{1, 2, \dots, m\} \} \subseteq F$.
		The acceptance of states of $Q_\top$ is defined similarly as for $Q_\bot$ except that the origins of vector spaces are accepting in $Q_\top$ (see Figure~\ref{fig:hard:example}).
		Formally, the rejecting states of $\A$ is defined by $\overline{F} = R_\bot \cup R_\top$ where $R_\bot = \{(q_1, q_2, q_3, 0) \in Q_\bot \st q_2 = vq_1 \mod \p, 1 \leq q_3 \leq m, v \in C_{q_3}\}$ and $R_\top = \{(q_1, q_2, q_3, 1) \in Q_\top \st (q_1,q_2,q_3,0) \in R_\bot, q_1\neq 0, q_2 \neq0\}$.
		All other states are accepting, i.e., we set $F= Q\setminus\overline{F}$.
		So, the acceptance of the subsets of states $Q_\bot$ and $Q_\top$ only differ by $O \cap Q_\bot \subseteq \overline{F}$ and $O \cap Q_\top \subseteq F$ where $O = \{ (0, 0, q_3, q_4) \in Q \st q_3 \in \{1, \dots, m\}\}$.
		
		%Formally, the \DFA $\A = \zug{\Sigma, Q, q_I, \delta, F}$ is defined over the alphabet $\Sigma = \{a, b, c, d\}$ where the transition function is defined for all state $q = (q_1, q_2, q_3, q_4)$ by $\delta(q, a) = (q_1+1 \mod \p,q_2,q_3,q_4)$, $\delta(q, b) = (q_1,q_2+1 \mod \p,q_3,q_4)$, $\delta(q, c) = (q_1,q_2,q_3+1 \mod \q,q_4)$ and $\delta(q, d) = (q_1,q_2,q_3,q_4+1\mod 2)$.
		%The initial state is $q_I = (0, 0, 0, 0)$ and the set of 
		%Formally, rejecting states is $\overline{F} = R_\bot \cup R_\top$ where $R_\bot = \{(q_1, q_2, q_3, 0) \in Q_\bot \st q_2 = vq_1 \mod \p, 1 \leq q_3 \leq m, v \in C_{q_3}\}$ and $R_\top = \{(q_1, q_2, q_3, 1) \in Q_\top \st (q_1,q_2,q_3,0) \in R_\bot, q_1\neq 0, q_2 \neq0\}$.
		%All other states are accepting, i.e., we set $F= Q\setminus\overline{F}$.
		The cornerstone which holds the connection between the two problems is the way the rejecting states of $O$ can be covered.
		In fact, since $Q_\top$ mimics $Q_\bot$ for states in $Q \setminus O$, all rejecting states of $Q \setminus O$ can be covered by the single word $d \in \Sigma$.
		In addition, most words do not cover any rejecting states of $\A$, as stated by the following claim.
		%\begin{restatable}{claim}{restatableClaimNotOrigin}\label{claim:not_origin}
		%	The word $d \in \Sigma$ covers exactly all rejecting states $\overline{F} \setminus O$.
		%\end{restatable}
		%\begin{toappendix}
		%	In the proof of Lemma~\ref{lemma:bounded_dec_hard}, we are using the property of $\A$ that every rejecting states that is not in $O$ can be covered by the word $d \in \Sigma$.
		%	\restatableClaimNotOrigin*
		%	\begin{proof}
		%	By construction $(q_1, q_2, q_3, 0) \in F \iff (q_1, q_2, q_3, 1) \in F$ holds for all $(q_1, q_2, q_3, q_4) \in Q \setminus O$.
		%	Thus, for all $(q_1, q_2, q_3, q_4) \in \overline{F} \setminus O$ we have $\{ \delta((q_1, q_2, q_3, q_4), d^\lambda) \st \lambda \in \N\} = \{(q_1, q_2, q_3, x) \st x \in \{0, 1\} \} \subseteq \overline{F}$.
		%	In addition, $\overline{F} \cap O = \{(0, 0, q_3, 0) \in Q_\bot \st 1\leq q_3 \leq m\}$ and $\{ \delta((0, 0, q_3, 0), d) \st 1 \leq q_3 \leq m\} \subseteq F$. %\hfill$\vartriangleleft$
		%	\end{proof}
		%\end{toappendix}
		%\begin{restatable}{claim}{restatableClaimBIS}\label{claim:all}
		%	Let $u$ be a word that covers some rejecting state of $\A$.
		%	\begin{enumerate}
		%		\item Either $u \in \{d\}^*$ and $u$ covers some rejecting states of $Q_\top$ iff $u=d$.
		%		\item Or $u \in \{a, b\}^*$ and $\#(u)_b \equiv v \cdot \#(u)_a \mod \p$ for some $v \in C_i$ iff $u$ covers $(0, 0, i, 0) \in O$.
		%	\end{enumerate}
		%\end{restatable}
		%The automaton $\A$ constructed in the proof of Lemma~\ref{lemma:bounded_dec_hard}, admits only restricted shapes of words that can actually cover some rejecting states as stated by the following Claim.
		Hereafter, we say that a word $w \in \Sigma^*$ is \emph{\shortish} when it satisfies $\#_\sigma(w) < h_\sigma$ for all $\sigma \in \Sigma$,
		where $h_\sigma \in \{2, \p, \q\}$ is the size of the cycle induced by $\sigma$.
		%
		%Let $h_a = \p, h_b = \p, h_c = \q, h_d=2$, which represent the sizes of each dimension of the set of states of $\A$.
		%A word $w \in \Sigma^*$ is said to be \shortish when it satisfies $\#(w)_\sigma < h_\sigma$ for all $\sigma \in \Sigma$.
		
		\begin{restatable}{claim}{restatableClaimAll}\label{claim:all}
			Let $u \in \Sigma^*$ be a \shortish word that covers some rejecting state of $\A$:
			%
			%	 that covers some rejecting state of $\A$ and have a run that does not loop
			%	\petra{1. Either $u \in {d}^*$ and $u$ covers some rejecting states of $Q_\top$ iff $u=d$.\\
			%		2. Or $u \in \{a, b\}^*$ and $\#(u)_b \equiv v \cdot \#(u)_a \mod \p$ for some $v \in C_i$ iff $u$ covers $(0, 0, i, 0) \in O$}
			\begin{enumerate}
				\item $u$ must belong either in $\{d\}^*$ or in $\{a, b\}^* \setminus (\{a\}^* \cup \{b\}^*)$.
				\item $u$ covers some rejecting state of $Q_\top$ iff $u$ covers \emph{all} rejecting states of $Q_\top$ iff $u = d$.
				\item $u$ covers $(0, 0, i, 0) \in O$ iff $u \in \{a, b\}^*$ and $\#_b(u) \equiv v \cdot \#_a(u) \mod \p$ for some $v \in C_i$.
			\end{enumerate}
		\end{restatable}
		
		%	\restatableClaimAll*
		
		\begin{proof}[Proof of Item 1.]
			The statement is a direct consequence of the following:
			\begin{description}
				\item[i.]
				Every \shortish word $u$ satisfying $\#_c(u)>0$ covers no rejecting state of $\A$;
				\item[ii.]
				Every \shortish word $u \in \{a\}^* \cup \{b\}^*$ covers no rejecting state of $\A$;
				\item[iii.]
				Every \shortish word $u$ satisfying $\#_a(u)>0$ and $\#_d(u)>0$ covers no rejecting state of $\A$;
				\item[iv.]
				Every \shortish word $u$ satisfying $\#_b(u)>0$ and $\#_d(u)>0$ covers no rejecting state of $\A$.
				%			\item[(\raisebox{1pt}{\scalebox{.6}{$\bigstar$}})]
				%			For all $\sigma \in \{c, d\}$, every useful \shortish words $u$
				%			in which $\sigma$ occurs at least once
				%%			satisfying $\#_\sigma(u) > 0$
				%%			that contains an occurrence of $\sigma$
				%			is in $\{\sigma\}^*$.
				%			\item[(\raisebox{.7pt}{\scalebox{.65}{$\blacklozenge$}})]
				%			For all $\sigma \in \{a, b, c\}$,
				%			there is no useful \shortish word in $\{\sigma\}^*$.
			\end{description}
			%		Applying (\raisebox{1pt}{\scalebox{.6}{$\bigstar$}}) with $\sigma = d$ allows us to show that any useful word cannot share occurrences of $d$ with occurrences of another letter of $\Sigma$.
			%		And similarly for $\sigma = c$.
			%		In addition, (\raisebox{.7pt}{\scalebox{.65}{$\blacklozenge$}}) allows us to show that
			%		no useful word  belongs to $\{a\}^* \cup \{b\}^* \cup \{c\}^*$.
			%		Therefore, every useful word must belongs to $\{d\}^* \cup \{a, b\}^* \setminus (\{a\}^* \cup \{b\}^*)$.
			In order to prove these four properties,
			we now fix a state $q = (q_1,q_2,q_3,q_4) \in Q$,
			and we show that, in each case, iterating a word of the corresponding form
			starting from $q$ will eventually lead to an accepting state:
			
			(i.) Let $u$ be a \shortish word satisfying $\#_c(u)>0$.
			Since $u$ is \shortish we have $\#_c(u)<\tau$.
			Hence, as $\tau$ is prime, there exists $\lambda \in \N$
			such that $\lambda \cdot \#_c(u) \equiv -q_3 \mod \tau$.
			Therefore the third component of $\delta(q,u^\lambda)$ is $0$,
			thus it is an accepting state of $\A$.
			
			(ii.) Let $u \in \{a\}^*$ be a \shortish word
			(if $u \in \{b\}^*$ instead, the same proof works by swapping the roles of $q_1$ and $q_2$).
			Since $u$ is \shortish we have $0<\#_a(u)<\mu$.
			Hence, as $\mu$ is prime there exists $\lambda_1,\lambda_2 \in \N$
			satisfying $\lambda_1 \cdot \#_a(u) \equiv -q_1 \mod \mu$ and $\lambda_2 \cdot \#_a(u) \equiv -q_1 +1 \mod \mu$.
			Therefore, if $q_2 \neq 0$, we get that $\delta(q,u^{\lambda_1}) = (0,q_2,q_3,q_4)$
			is an accepting state of $\A$,
			and if $q_2 = 0$, we get that $\delta(q,u^{\lambda_2}) = (1,0,q_3,q_4)$
			is an accepting state of $\A$.
			
			(iii.) Let $u$ be a \shortish word satisfying $\#_a(u)>0$ and $\#_d(u)>0$.
			Since $\mu$ is a prime number greater than $2$,
			there exist $\alpha \in \N$
			such that $\mu - 2 \alpha = 1$, thus $2 \alpha \equiv -1 \mod \mu$.
			Moreover, since $u$ is \shortish we have $\#_d(u) = 1$ and $\#_a(u)<\mu$.
			Hence there exists $\beta \in \N$
			such that $\beta \cdot \#_a(u) \equiv 1 \mod \mu$.
			Therefore, if we let $\lambda = 2 \alpha \beta q_1 + \mu(1-p_4)$, we get
			\begin{align*}
				&\#_{a}(u^{\lambda})
				%			=(2 \alpha \beta q_1 + \mu(1-p_4)) \cdot \beta\#_{a}(u)
				= 2 \alpha \cdot \beta\#_{a}(u) \cdot q_1  +  \mu (1-p_4) \cdot  \#_{a}(u) 
				\equiv - q_1 \mod \mu;\\
				&\#_{d}(u^{\lambda})
				%			= (2 \alpha \beta q_1 + \mu(1-p_4)) \cdot \beta\#_{d}(u)
				= 2 \alpha \beta q_1 + \mu \cdot (1-p_4)
				\equiv 1-p_4 \mod 2;
			\end{align*}
			As a consequence, the first component of $\delta(q,u^{\lambda})$
			is $0$ and its fourth component is $1$,
			hence it is an accepting state of $\A$.
			
			(iv.) Let $u$ be a \shortish word satisfying $\#_b(u)>0$ and $\#_d(u)>0$.
			Then we can prove that $u$ does not cover $q$
			as in point (3), by swapping the roles of $q_1$ and $q_2$.
		\end{proof}
		
		\begin{proof}[Proof of Item 2.]
			First, remark that $d$ is the only \shortish word of $\{d\}^*$.
			By construction of $\A$, we have $(q_1, q_2, q_3, 0) \in F$ if and only if $(q_1, q_2, q_3, 1) \in F$
			holds for all $(q_1, q_2, q_3, q_4) \in Q \setminus O$.
			Thus, for all $(q_1, q_2, q_3, q_4) \in \overline{F} \setminus O$ we have
			\[
			\{ \delta((q_1, q_2, q_3, q_4), d^\lambda) \st \lambda \in \N\} = \{(q_1, q_2, q_3, x) \st x \in \{0, 1\} \} \subseteq \overline{F}.
			\]
			%In addition, $\overline{F} \cap O = \{(0, 0, q_3, 0) \in Q_\bot \st 1\leq q_3 \leq m\}$ and $\{ \delta((0, 0, q_3, 0), d) \st 1 \leq q_3 \leq m\} \subseteq F$.
			Hence, if $u = d$ then $u$ covers all rejecting states of of $Q_\top$.
			
			Now suppose that $u \in \Sigma^*$ covers some rejecting state $q = (q_1,q_2,q_3,1) \in Q_{\top}$.
			By Item~(1.),
			either $u \in \{d\}^*$ or $u \in \{a, b\}^* \setminus (\{a\}^* \cup \{b\}^*)$.
			We show that $u \in \{d\}^*$, by supposing that $\#_{a}(u)>0$ and deriving a contradiction.
			Since $\mu$ is prime, 
			there exists $\lambda \in \N$
			satisfying $\lambda \cdot \#_a(u) \equiv -q_1 \mod \mu$.
			Therefore the first component of $\delta(q,u^{\lambda})$ is $0$ and its fourth component is $1$,
			hence it is accepting, which contradicts the assumption that $u$ covers $q$.\qedhere
		\end{proof}
		
		\begin{proof}[Proof of Item 3.]
			Consider a rejecting state $q = (0,0,i,0) \in O$.
			First, remark that no word in $\{d\}^*$ covers $q$ since $(0,0,i,1)$ is accepting.
			Therefore, by Item~(1.),
			the only concise words that can cover $q$ are the words $u \in \{a,b\}^* \setminus (\{a\}^* \cup \{b\}^*)$.
			For such a word $u$, since $\mu$ is prime,
			by Bezout's identity there exists $0 < v < \mu$
			satisfying $\#_b(x) \equiv v \cdot \alpha \#_a(x) \mod \mu$,
			hence 
			\[
			\{ \delta((0, 0, i, 0), u^\lambda) \st \lambda \in \N \} =
			\{ (q_1, q_2, i, 0) \in Q \st q_2 \equiv vq_1 \mod \p \}.
			\]
			If $v \in C_i$, all the states in this set are rejecting, thus $u$ covers $(0,0,i,0)$,
			but if $v \notin C_i$, all these states except from $(0,0,i,0)$ are accepting, thus $u$ does not cover $(0,0,i,0)$.
			\qedhere
			% % % % % % % % PREVIOUS PROOF % % % % % % % %
			%Remark that, we can assume that $\#(u)_a > 0$ and $\#(u)_b > 0$ otherwise $u$ does not cover any rejecting states of $\A$ by Claim~\ref{claim:one}.
			%For all $1 \leq \lambda < \p$ we have $\#(u^\lambda)_a \not \equiv 0 \mod \p$ since $\p$ is co-prime with both $\#(u)_a$ and $\lambda$.
			%We recall that $\delta(q, a^{0}) = \delta(q, a^{\p}) = q$ for all $q \in Q$.
			%As a consequence, for all $x\in \N$ there exists $1 \leq \lambda \leq \p$ such that $\#(u^\lambda)_{a} \equiv x \mod \p$.
			%The statement (\textcolor{lipicsGray}{\sffamily\bfseries 1.}) holds since $\delta(q, u^{\lambda}) \in F$ for all $q \in Q_\top$ when $\lambda$ fulfills $\#(u^{\lambda})_{a} \equiv 0 \mod \p$.
			%To prove (\textcolor{lipicsGray}{\sffamily\bfseries 2.}) we define, for all $1 \leq i \leq m$ and all $v \in C_i$, the natural $\lambda_{i,v} \in \N$ such that $\#(u^{\lambda_{i,v}})_{a} \equiv 1 \mod \p$.
			%By contrapositive, if $\forall v \in C_i,\,\#(u)_b \not\equiv v\cdot\#(u)_a \mod \p$ then $\forall v\in C_i,\,\#(u^{\lambda_{i,v}})_{b} \not\equiv v\cdot\#(u^{\lambda_{i,v}})_{a} \equiv v \mod \p$ and thus $\forall v\in C_i,\, \delta(q, u^{\lambda_{i,v}}) \in F$ for all $q \in O$.
			%The other direction is trivial from the construction of $\A$, as a matter of fact, 
			%for all $1 \leq i \leq m$ and all $v \in C_i$ we have $\{ \delta((0, 0, i, 0), u^\lambda) \st \lambda \in \N \} = \{ (q_1, q_2, i, 0) \in Q \st q_2=vq_1 \mod \p \} \in \overline{F}$.
			% % % % % % % %  % % % % % % % %  % % % % % % % %
		\end{proof}

		We finally conclude the proof of Lemma \ref{lemma:bounded_dec_hard}
		by proving that the sets of the initial instance of \HIT are hitting if and only if the automaton $\A$ is composite.
		
		\boldtitle{If sets are hitting then the automaton is composite}
		Thanks to Lemma~\ref{lemma:Newword_dec=cover}, 
		we can show that $\A$ is $(k+1)$-factor composite by finding $(k+1)$ words, namely $w_\top, w_1, w_2, \dots, w_{k}$, which all together cover all the rejecting states of $\A$.
		From the \HIT solution $X =\{v_1, v_2, \dots, v_k\} \subseteq S$, we define $w_j = ab^{v_j}$ for all $1 \leq j \leq k$.
		We prove now that for all $1 \leq i \leq m$, the rejecting state $(0, 0, i, 0) \in O$ is covered by some~$w_j$.
		Since $X \cap C_i \neq \varnothing$, there exists $v_j \in X \cap C_i$.
		Moreover, by definition of $w_j$,
		we have $w_j \in \{a, b\}^*$ and $\#_b(w_j) \equiv v_j \cdot \#_a(w_j) \mod \p$.
		Therefore, by Claim~\ref{claim:all}.3,
		$(0,0,i,0)$ is covered by $w_j$.
		Finally, we take $w_\top = d$ which covers all rejecting states $\overline{F} \setminus O$ by Claim~\ref{claim:all}.2.

		\boldtitle{If the automaton is composite then the sets are hitting}
		Suppose that $\A$ is $(k+1)$-factor composite.
		Hence, by Lemma~\ref{lemma:Newword_dec=cover}, there exists a set $W$ of at most $k+1$ words
		%, namely $W = \{w_\top, w_1, \dots, w_{k}\}$,
		such that all rejecting states of $\A$ can be covered by some $w \in W$.
		In addition, we assume that each $w \in W$ is \shortish: if this is not the case,
		we can remove the superfluous letter to obtain a \shortish words that cover the same rejecting states.
		%The intuition is that most \shortish words do not cover any rejecting states, which in turn, allows us to consider only specific ones.
		As a consequence of Claim~\ref{claim:all}.2,
		to cover the rejecting states of $Q_\top$,
		the set $W$ needs the word $d$,
		thus $W$ contains at most $k$ words in $\{a,b\}^*$.
		Moreover, by Claim~\ref{claim:all}.3,
		for every $1 \leq i \leq m$, to cover $(0,0,i,0) \in O$
		the set $W$ needs a word
		$u_i \in \{a, b\}^*$ satisfying $\#_b(u_i) \equiv v_i \cdot \#_a(u_i) \mod \p$ for some $v_i \in C_i$.
		% we can assume w.l.o.g that $W\cap \{d\}^* \neq \varnothing$.
		%Let $w_\top\in \{d\}^*$ and $1 \leq |w_\top| \leq 2$, thus $w_\top = d$ and covers all rejecting states of $Q_\top$ by Claim~\ref{claim:all}.2.
		To conclude, we construct $X = \{ v_i \st 1 \leq i \leq m\}$ which is a solution since $|X| \leq k$ due to $W\cap \{d\}^* \neq \varnothing$,
		and for each $C \in \mathcal{F}$ we have $X\cap C \neq \varnothing$.
		% by Claim~\ref{claim:all}.3.
		\qedhere
	\end{proof}

	\section{Discussion}
	We introduced in this work powerful techniques to treat
	the \Decomp problem for permutation \DFAs.
	We discuss how they could help solving the
	related questions that remain open:
	\begin{itemize}
		\item How do the insights obtained by our results translate to the general setting?
		\item How can we use our techniques to treat other variants of the \Decomp problem? 
	\end{itemize}
	
	\boldtitle{Solving the general setting}
	The techniques presented in this paper rely heavily on the group structure of
	transition monoids of permutation \DFAs,
	thus cannot be used directly in the general setting.
	They still raise interesting questions:
	Can we also obtain an \FPT algorithm with respect to the number of rejecting states
	in the general setting?
	Some known results point that bounding the number of states is not as useful
	in general as it is for permutation \DFAs:
	while it is known that every permutation \DFA with a single rejecting state is prime~\cite{KM15},
	there exist (non-permutation) \DFAs with a single rejecting state that are composite.
	However, we still have hope to find a way to adapt our techniques:
	maybe, instead of trying to cover rejecting \emph{states},
	we need to cover rejecting \emph{behaviours} of the transition monoid.
	Another way to improve the complexity in the general setting
	would be to bound the \emph{width} of \DFAs:
	we defined here a family of \DFAs with polynomial width,
	do there exist families with exponential width?
	If this is not the case (i.e., every composite \DFA has polynomial width),
	we would immediately obtain a \PSPACE algorithm for the general setting.

	\boldtitle{Variants of the \Decomp problem}
	In this work, we focused on the \boundDecomp problem,
	that limits the \emph{number} of factors in the decompositions.
	Numerous other restrictions can be considered.
	For instance, the \textsc{Fragmentation} problem bounds the \emph{size} of the factors:
	Given a \DFA $\A$ and $k \in \mathbb{N}$,
	can we decompose $\A$ into \DFAs of size smaller than $k$?
	Another interesting restriction is proposed by the \textsc{Compression} problem,
	that proposes a trade-off between limiting the size and the number of the factors:
	given a \DFA $\A$, can we decompose $\A$ into \DFAs
	$(\A_i)_{1 \leq i \leq k}$ satisfying $\Sigma_{i=1}^n |\A_i| < |\A|$?
	How do these problems compare to the ones we studied?
	We currently conjecture that the complexity of the \textsc{Fragmentation} problem matches the \Decomp problem,
	while the complexity of the \textsc{Compression} problem matches the \boundDecomp problem:
	for commutative permutation \DFAs, the complexity seems to spike
	precisely when we limit the number of factors.
	
	%Another interesting direction, would be to consider a tradeoff between the number of factors and they sizes.
	%Such approach have been investigated in \cite{netser}, where the author provides a family of composite \DFAs of size $O(n)$ for which there exists a decomposition with $O(\sqrt{n})$ factors of size $O(\sqrt{n})$.
	%As a natural variate of the compositionality, we define now the compression of a composite \DFA $\A$ by a decomposition $\A_1, \A_2, \dots, \A_k$ such that $L(\A) = \bigcap_{i=1}^k L(\A_i)$ and $\sum_{i=1}^{k} |\A_i| < |\A|$.
	%In particular, any compressions have at most $|\A|$ factors.
	%As a matter of fact, we conjecture that our proof of Section~\ref{sec:bounded_decomposition} can be adapted for this problem.
	%More precisely, containing solutions of the \textsc{Hitting Set} problem of be of size at most half the total number of elements (i.e. $k < n/2$) allows us to show the \NP-hardness for compression.
	%EASINESS?
	
	\bibliography{ok}{}

\begin{thebibliography}{10}

\bibitem{DBLP:books/daglib/0020348}
Christel Baier and Joost{-}Pieter Katoen.
\newblock {\em Principles of Model Checking}.
\newblock {MIT} Press, 2008.

\bibitem{clarke1991language}
Edmund~M. Clarke, David~E. Long, and Kenneth~L. McMillan.
\newblock A language for compositional specification and verification of finite
  state hardware controllers.
\newblock {\em Proceedings of the IEEE}, 79(9):1283--1292, 1991.
\newblock \href {https://doi.org/10.1109/5.97298} {\path{doi:10.1109/5.97298}}.

\bibitem{dRLP98}
Willem~P. de~Roever, Hans Langmaack, and Amir Pnueli, editors.
\newblock {\em Compositionality: The Significant Difference, International
  Symposium, COMPOS'97, Bad Malente, Germany, September 8-12, 1997. Revised
  Lectures}, volume 1536 of {\em Lecture Notes in Computer Science}. Springer,
  1998.
\newblock \href {https://doi.org/10.1007/3-540-49213-5}
  {\path{doi:10.1007/3-540-49213-5}}.

\bibitem{10.5555/578533}
Michael~R. Garey and David~S. Johnson.
\newblock {\em Computers and Intractability: A Guide to the Theory of
  NP-Completeness}.
\newblock W. H. Freeman \& Co., USA, 1979.

\bibitem{gould2007apparatus}
Stephen Gould, Ernest Peltzer, Robert~Matthew Barrie, Michael Flanagan, and
  Darren Williams.
\newblock Apparatus and method for large hardware finite state machine with
  embedded equivalence classes, 2007.
\newblock US Patent 7,180,328.

\bibitem{hardy1929}
G.~H. Hardy.
\newblock An introduction to the theory of numbers.
\newblock {\em Bulletin of the American Mathematical Society}, 35(6):778--818,
  11 1929.
\newblock URL: \url{https://projecteuclid.org:443/euclid.bams/1183493592}.

\bibitem{JKM20}
Isma{\"{e}}l Jecker, Orna Kupferman, and Nicolas Mazzocchi.
\newblock Unary prime languages.
\newblock In Javier Esparza and Daniel Kr{\'{a}}l, editors, {\em 45th
  International Symposium on Mathematical Foundations of Computer Science,
  {MFCS} 2020, August 24-28, 2020, Prague, Czech Republic}, volume 170 of {\em
  LIPIcs}, pages 51:1--51:12. Schloss Dagstuhl - Leibniz-Zentrum f{\"{u}}r
  Informatik, 2020.
\newblock \href {https://doi.org/10.4230/LIPIcs.MFCS.2020.51}
  {\path{doi:10.4230/LIPIcs.MFCS.2020.51}}.

\bibitem{DBLP:conf/mfcs/KuncO13}
Michal Kunc and Alexander Okhotin.
\newblock Reversibility of computations in graph-walking automata.
\newblock In Krishnendu Chatterjee and Jir{\'{\i}} Sgall, editors, {\em
  Mathematical Foundations of Computer Science 2013 - 38th International
  Symposium, {MFCS} 2013, Klosterneuburg, Austria, August 26-30, 2013.
  Proceedings}, volume 8087 of {\em Lecture Notes in Computer Science}, pages
  595--606. Springer, 2013.
\newblock \href {https://doi.org/10.1007/978-3-642-40313-2\_53}
  {\path{doi:10.1007/978-3-642-40313-2\_53}}.

\bibitem{KM15}
Orna Kupferman and Jonathan Mosheiff.
\newblock Prime languages.
\newblock {\em Inf. Comput.}, 240:90--107, 2015.
\newblock \href {https://doi.org/10.1016/j.ic.2014.09.010}
  {\path{doi:10.1016/j.ic.2014.09.010}}.

\bibitem{DBLP:journals/ibmrd/Landauer61}
Rolf Landauer.
\newblock Irreversibility and heat generation in the computing process.
\newblock {\em {IBM} J. Res. Dev.}, 5(3):183--191, 1961.
\newblock \href {https://doi.org/10.1147/rd.53.0183}
  {\path{doi:10.1147/rd.53.0183}}.

\bibitem{doi:10.4169/amer.math.monthly.120.07.650}
Jaban Meher and M~Ram Murty.
\newblock Ramanujan's proof of {B}ertrand's postulate.
\newblock {\em The American Mathematical Monthly}, 120(7):650--653, 2013.
\newblock URL:
  \url{https://www.tandfonline.com/doi/abs/10.4169/amer.math.monthly.120.07.650},
  \href {https://doi.org/10.4169/amer.math.monthly.120.07.650}
  {\path{doi:10.4169/amer.math.monthly.120.07.650}}.

\bibitem{netser}
Alon Netser.
\newblock Decomposition of safe languages.
\newblock Amirim Research Project report from the Hebrew University, 2018.

\bibitem{pedroni2013finite}
Volnei~A. Pedroni.
\newblock {\em Finite State Machines in Hardware: Theory and Design (with VHDL
  and SystemVerilog)}.
\newblock The MIT Press, 2013.

\bibitem{DBLP:conf/latin/Pin92}
Jean{-}Eric Pin.
\newblock On reversible automata.
\newblock In Imre Simon, editor, {\em {LATIN} '92, 1st Latin American Symposium
  on Theoretical Informatics, S{\~{a}}o Paulo, Brazil, April 6-10, 1992,
  Proceedings}, volume 583 of {\em Lecture Notes in Computer Science}, pages
  401--416. Springer, 1992.
\newblock \href {https://doi.org/10.1007/BFb0023844}
  {\path{doi:10.1007/BFb0023844}}.

\end{thebibliography}
	
\end{document}